\documentclass[sigconf,balance=false]{acmart}
\usepackage{popets}

\usepackage{nicefrac}
\usepackage{siunitx}
\usepackage{array,framed}
\usepackage{booktabs}
\usepackage{
  color,
  float,
  epsfig,
  wrapfig,
  graphics,
  graphicx,
  subcaption
}
\usepackage{amsthm}
\usepackage{textcomp}
\usepackage{setspace}
\usepackage{latexsym,fancyhdr,url}
\usepackage{enumerate}
\usepackage[linesnumbered,ruled]{algorithm2e}
\usepackage{algpseudocode}
\usepackage{graphics}
\usepackage{xparse} % argument parsing -- \edist
\usepackage{xspace}
\usepackage{multirow}
\usepackage{csvsimple}
\usepackage{balance}
\usepackage{subcaption}
\usepackage{multirow}

\newtheorem*{clm}{Claim}
\setcopyright{popets}
\copyrightyear{YYYY}

\acmYear{YYYY}
\acmVolume{YYYY}
\acmNumber{X}
\acmDOI{XXXXXXX.XXXXXXX}
\acmISBN{}
\acmConference{Proceedings on Privacy Enhancing Technologies}
\begin{document}

\title{Robust Fingerprint of Location Trajectories Under Differential Privacy}

%%%%%%%%%%%%%%%% Authors' Info %%%%%%%%%%%%%%%%%
\author{Yuzhou Jiang}
\affiliation{%
  \institution{Case Western Reserve University}
  \city{Cleveland}
  \state{Ohio}
  \country{USA}}
\email{yxj466@case.edu}

\author{Emre Yilmaz}
\affiliation{%
  \institution{University of Houston-Downtown}
  \city{Houston}
  \state{Texas}
  \country{USA}}
\email{yilmaze@uhd.edu}

\author{Erman Ayday}
\affiliation{%
  \institution{Case Western Reserve University}
  \city{Cleveland}
  \state{Ohio}
  \country{USA}}
\email{exa208@case.edu}


\renewcommand{\shortauthors}{Jiang et al.}

\begin{abstract}
Location-based services have brought significant convenience to people in their daily lives, and the collected location data are also in high demand. 
However, directly releasing those data raises privacy and liability (e.g., due to unauthorized distribution of such datasets) concerns since location data contain users' sensitive information, e.g., regular moving patterns and favorite spots. To address this, we propose a novel fingerprinting scheme that simultaneously identifies unauthorized redistribution of location datasets and provides differential privacy guarantees for the shared data. 
Observing data utility degradation due to differentially-private mechanisms, we introduce a utility-focused post-processing scheme to regain spatio-temporal correlations between points in a location trajectory. We further integrate this post-processing scheme into our fingerprinting scheme as a sampling method.
The proposed fingerprinting scheme alleviates the degradation in the utility of the shared dataset due to the noise introduced by differentially-private mechanisms (i.e., adds the fingerprint by preserving the publicly known statistics of the data).
Meanwhile, it does not violate differential privacy throughout the entire process due to immunity to post-processing, a fundamental property of differential privacy.
Our proposed fingerprinting scheme is robust against known and well-studied attacks against a fingerprinting scheme including random flipping attacks, correlation-based flipping attacks, and collusions among multiple parties, which makes it hard for the attackers to infer the fingerprint codes and avoid accusation. Via experiments on two real-life location datasets and two synthetic ones, we show that our scheme achieves high fingerprinting robustness and outperforms existing approaches. Besides, the proposed fingerprinting scheme increases data utility for differentially-private datasets, which is beneficial for data analyzers.

\end{abstract}


\maketitle
\vspace{-2mm}
\section{Introduction}
\label{sec:introduction}

Location-based services have become one of the most popular services in our daily lives thanks to rapid evolution in mobile technologies and internet of things.
Location-based service providers often require a large amount of location-based information from users to support their services.
For instance, Google Maps~\cite{gomaps_web} collects users' accurate location data in real-time and plans optimal routes during navigation and offers place suggestions while users are searching on the app. Food delivery services, e.g., Doordash~\cite{doordash_web}, demand approximate location information from users for restaurant recommendation and keep track of food couriers for better user experience. Most individuals are subtly accustomed to the convenient lifestyles using these location-based services, and hence they share their location data with such location-based service providers voluntarily (with consent). Thus, such service providers build large location datasets.

Location datasets are of great use, and sharing them bring vast benefits. Besides moving patterns, much more information are included and can be inferred from these datasets (e.g., age, job, or home address). By analyzing the datasets, data analytics companies can offer proper suggestions to the service providers in order to improve their user experience, adjust marketing strategies, or even determine locations of new facilities. Advertisement companies can learn from those data for accurate promotion to specific customers. Researchers can propose new approaches and validate them on these datasets. 

Location-based service providers (e.g., Google) can share such location datasets with a limited number of parties, called data analyzers. Some examples of data analyzers are researchers and analytic institutions. Access to location datasets are typically restricted within such analyzers parties as location datasets contain sensitive information. 
Nevertheless, malicious data analyzers, e.g., motivated by profit, may leak their copies to unauthorized parties, which brings significant privacy concerns. In order to prevent unauthorized redistribution, service providers should embed a unique fingerprint into datasets for each data analyzer to enable traceability of the potential leakage. Such fingerprint should be robust against multiple attacks, e.g., distortion attacks and collusion attacks, since the attackers may try distort it by modifying some points or even colluding with other malicious parties to get rid of accusation.
By analyzing the embedded fingerprint in the leaked dataset, the service provider can identify the source of the leakage, withdraw its access to the dataset, and even punish it. Thus, knowing that the leaked dataset will be traced back to them, attackers become less motivated to leak the copies of received datasets. 

There are several existing fingerprinting mechanisms, e.g., Boneh-Shaw codes~\cite{boneh1998collusion} and Tardos codes~\cite{tardos2008optimal}. However, those traditional digital fingerprinting schemes cannot be directly applied to the location datasets because of correlations in location datasets and their particular utility requirements. In a location trajectory, i.e., an ordered sequence of location points in a location dataset, location points are highly correlated with each other, especially the adjacent location points. 
For instance, in a walking trajectory, recorded at every 10 seconds, it is not likely to have two contiguous location points one kilometer apart from each other. 
Also, by knowing the previous and the following points in a given location trajectory, one can precisely estimate/infer the intermediate point with high confidence. 
Thus, using publicly available correlation models (constructed from public location datasets), an attacker can identify the points that violate the expected correlations as the fingerprinted data points. 
It can then distort or remove such identified data points (i.e., distort the fingerprint), which makes it harder for the service providers to detect the source of a leaked dataset. 
We observed (and show via experiments) that existing fingerprint codes are vulnerable to such correlation-based attacks since they do not consider pairwise correlations. 
Therefore, in this paper, we propose a robust correlation-based fingerprinting scheme that is robust against multiple attacks, e.g., including correlation attacks, majority collusion attacks, and probabilistic collusion attacks.

On the other hand, in recent years, privacy concerns of sensitive datasets have attracted massive attention. 
Researchers have also been investigating privacy of location data and location datasets~\cite{andres2013geo, de2013unique}. It has been shown that users' identities can be deanonymized with high confidence given only a pattern of four location points \cite{de2013unique}. Therefore, simple anonymization on identifiers/quasi-identifiers is not sufficient to protect the individuals' location privacy. 
Under differential privacy (DP - a state-of-the-art concept for privacy preservation that quantifies and limits the information acquired from the attackers' perspective), researchers have proposed several solutions to mitigate privacy leakage while sharing location data, e.g., PIM~\cite{xiao2015protecting} and AdaTrace~\cite{gursoy2018utility}. 
However, existing privacy preserving approaches for location data and datasets (i) do not provide liability guarantees against dataset leakage (unauthorized redistribution); and (ii) bring excessive noise to datasets and thus sacrifice data utility. 
Some location-based services (e.g., navigation) that do not tolerate such low utility may be unwilling to apply privacy protection to their datasets.

To the best of our knowledge, no existing work can tackle both issues, i.e., guaranteeing differential privacy and offering fingerprinting robustness, simultaneously. It is true that one can apply an arbitrary differentially private mechanism followed by an existing fingerprinting scheme, or vice versa. However, such differentially private mechanisms or fingerprinting schemes have their own drawbacks for location datasets. 
For instance, existing methods that achieve differential privacy on location datasets either omit critical information~\cite{gursoy2018utility} or they require impractical restrictions~\cite{jiang2013publishing}. In terms of fingerprinting, existing schemes~\cite{boneh1998collusion, ji2021curse, tardos2008optimal} are limited in their ability to account for correlations in location datasets. These schemes often require specific types of data and they do not incorporate such correlations in their methodology, thus resulting in significant utility loss in the shared dataset. 
To solve these problems, we propose our solution that ensures differential privacy guarantee and high fingerprinting robustness along with high data utility at the same time.

In this work, we introduce a robust fingerprinting scheme for location datasets that are protected under differential privacy using probabilistic sampling. The proposed scheme checks spatial and temporal correlations along the trajectories and considers highly probable location points based on public correlations during fingerprinting. The fingerprinting scheme offers high detection accuracy against multiple attacks against a fingerprinting scheme, e.g., random flipping attacks, correlation-based flipping attacks, majority collusion attacks, and probabilistic collusion attacks~\cite{yilmaz2020collusion}.
The selection of the privacy-preserving technique can be arbitrary. We select the planar isotropic mechanism (PIM)~\cite{xiao2015protecting} as the building block to achieve differential privacy.
Other differentially private approaches can be used as well (e.g.,  AdaTrace~\cite{gursoy2018utility}, a state-of-the-art synthetic approach for releasing location datasets under differential privacy). We demonstrate this flexibility of the proposed scheme through evaluations in Section~\ref{sec:eval_alt}.
To mitigate data utility degradation due to the privacy-preserving methods, we propose a utility-focused post-processing scheme that aims to restore correlations between adjacent points along a trajectory. During this process, we check the $2$-gram transitions in the trajectory and replace each location point that has a low probability with a highly probable one by considering the directional information of the transition. We integrate this post-processing scheme into our proposed fingerprinting scheme such that the fingerprinting scheme can protect unauthorized redistribution and boost data utility at the same time.

We implement our proposed scheme using two real-life datasets, i.e., the GeoLife dataset~\cite{zheng2010geolife} and the Taxi dataset~\cite{moreira2013predicting}, and two synthetic datasets generated from Brinkhoff generator~\cite{brinkhoff2002framework}. We compare our scheme with state-of-the-art fingerprinting approaches, i.e., Boneh-Shaw codes and Tardos codes, and evaluate the fingerprint robustness against random flipping attacks, correlation-based flipping attacks, majority collusion attacks, and probabilistic collusion attacks.
We also evaluate data utility in terms of query answering of location points and patterns, area popularity, trip error, diameter error, and trajectory similarity. We observe that our scheme provides significantly better data utility than the existing approaches.

Our main contributions can be summarized as follows:

\begin{itemize}
  \item We propose a probabilistic fingerprinting scheme that utilizes publicly known correlations for location datasets.
  \item We propose a utility-focused post-processing scheme to improve data utility for the location datasets that are protected under differential privacy and further integrate it into the proposed fingerprinting scheme.
  \item The fingerprinting scheme achieves high fingerprint robustness on differentially private datasets against several known attacks.
  \item We evaluate our proposed scheme concerning fingerprint robustness and data utility on four datasets, and show that our scheme outperforms state-of-the-art approaches.
\end{itemize}

The rest of the paper is organized as follows. We review the existing works in Section~\ref{sec:related_works} and provide the preliminaries in Section~\ref{sec:preliminaries}. We present the system and threat models in Section~\ref{sec:problems}. In Section~\ref{sec:methodology}, we introduce the proposed scheme in detail. We evaluate our proposed scheme in Section~\ref{sec:evaluation}. In Section~\ref{sec:disc}, we discuss several topics  related to our approach. Section~\ref{sec:conclusion} concludes the paper.

\vspace{-2.5mm}
\section{Related Work}
\label{sec:related_works}
In this section, we introduce some existing works in location privacy and digital fingerprinting, respectively.
\vspace{-3mm}
\subsection{Location Privacy}
\label{sec:related_privacy}
\vspace{-0.5mm}
Location data contain sensitive information such as moving patterns and preferred locations. 
Traditional privacy enhancing techniques, e.g., k-anonymity~\cite{sweeney2002k} and l-diversity~\cite{machanavajjhala2007diversity}, have been adapted to the location setting. However, for a location dataset, those techniques have their limitations in dealing with data streams with various lengths. For instance, some works~\cite{fung2008anonymity, abul2008never} split the trajectories into equal-length fragments and achieve privacy on the fragment, which is not sufficient for privacy protection on trajectories. Differential privacy~\cite{dwork2008differential} as a popular privacy definition has been used to protect location datasets in recent years~\cite{shokri2014privacy, yu2017dynamic, cao2017quantifying, shokri2011quantifying}. Geo-indistinguishability~\cite{andres2013geo} defines a variant of differential privacy based on the distance between the points of interests, but it only 
works on location points instead of trajectories. Several methods~\cite{chen2012differentially,gursoy2018utility, he2015dpt} provide differential privacy to the statistics from original location datasets. He et al~\cite{he2015dpt} design a hierarchical tree for storing regional spatial correlations and sample trajectories by walking along the tree paths.
Gursoy et al.~\cite{gursoy2018utility} extract four statistical features from a location dataset under differential privacy and generate a synthetic dataset using those noisy features. 
These works completely eliminate moving features of any specific user while preserving statistics, which improves user's location privacy but significantly decreases the usability of the dataset in certain services, e.g., map navigation and carpooling. Meanwhile, some researchers use perturbation-based approaches instead. \cite{jiang2013publishing} releases differentially private trajectories by sampling and interpolating them, but the scheme has an additional restriction that starting and ending locations should be known to the public. PIM~\cite{xiao2015protecting} distorts each location point in a trajectory based on prior knowledge from previously released points. This approach is the only existing one that takes spatio-temporal correlations into consideration during differentially private release. However, it introduces zig-zag patterns for lower privacy budgets (i.e., privacy protection is stronger) in the shared trajectories and loses pairwise correlation along a trajectory, making it also suffers from utility loss.

\vspace{-2mm}
\subsection{Digital Fingerprinting}
\label{sec:related_fingerprinting}
\vspace{-0.5mm}
Digital fingerprinting embeds a unique identifier, e.g., a sequence of marks, to the data by adding, removing or editing partial values of the data. Several works have been proposed to enable digital fingerprinting for data distribution~\cite{cheng2011anti, boneh1998collusion, tardos2008optimal}. Boneh and Shaw design a fingerprint code and prevent the receivers from colluding~\cite{boneh1998collusion}. Tardos et al. propose a probability-based fingerprinting scheme that can catch all suspicious individuals simultaneously~\cite{tardos2008optimal} and has less code length than Boneh and Shaw's. Wu et al. introduce a fingerprinting scheme that embeds binary fingerprint codes towards multimedia~\cite{wu2004collusion}.
However, those methods are designed for binary streams, where pairwise correlations are omitted in most cases.  Considering correlations, some researchers aim to provide fingerprint robustness in the data with various types, i.e., relational databases~\cite{li2005fingerprinting, ji2021curse, liu2004block, lafaye2008watermill}. These approaches only work on specific data types and cannot be applied to location datasets since location trajectories have high pairwise correlations. Considering correlations, \cite{yilmaz2020collusion} introduces a fingerprinting scheme for sequential data that considers correlations between data points. Still, it requires the possible states for a data point be limited, discrete, and inter-transitable.

\vspace{-2mm}
\section{Preliminaries}
\label{sec:preliminaries}

In this section, we first introduce the definition of differential privacy and its key property: immunity to post-processing. We then introduce two popular collusion-resistant fingerprinting schemes as the baseline approaches against collusion attacks. We integrate one of the schemes into our proposed robust fingerprinting scheme (i.e., the Boneh-Shaw codes) and compare it with the vanilla versions of these schemes in Section~\ref{sec:evaluation}.

\vspace{-2mm}
\subsection{Differential Privacy}
\label{sec:dp}
Differential privacy (DP) quantifies privacy and limits the inference of any single individual from observing the query results between neighboring databases. The formal definition is as follows:

\begin{definition}[Differential Privacy]~\cite{dwork2008differential}
\label{def:dp}
 For any neighboring datasets $D, D'$ that only differ in one data record, a randomized algorithm $\mathcal{M}$ satisfies $\epsilon$-differential privacy if for all possible outputs $\mathcal{S} \subseteq Range(\mathcal{M})$
$$ Pr(\mathcal{M}(D) \in \mathcal{S}) \leq e^\epsilon * Pr(\mathcal{M}(D') \in \mathcal{S})\text{.}$$
\end{definition}

An important proposition of differential privacy is its immunity to post-processing. It ensures that the differential privacy guarantee still holds when a mapping function is performed on the output from a differentially private mechanism as long as the function does not utilize the actual value. The formal definition is as follows: 
\begin{proposition}[Post-processing]~\cite{dwork2014algorithmic}
\label{prop:immunity}
Let $\mathcal{M}$ be a randomized algorithm that is $\epsilon$-differentially private. For any arbitrary randomized mapping $f:\mathcal{R}^q \rightarrow \mathcal{R}^r$ where $p,q \in \mathbb{N}^+$, $f\circ \mathcal{M}$ is $\epsilon$-differentially private.
\vspace{-1.5mm}
\end{proposition}

Hence, perturbations to the differentially private outputs without knowing the original values do not violate the privacy guarantee.

\vspace{-2mm}
\subsection{Planar Isotropic Mechanism}
\label{sec:pim}

The planar isotropic mechanism (PIM)~\cite{xiao2015protecting} aims to protect each location point along an individual's location trajectory under differential privacy. 
It constructs the correlations of a trajectory using a Markov chain, which is treated as a hidden Markov model from the attacker's perspective. Based on the adversarial knowledge, i.e., the probability distribution of the location, the method adds calibrated noise to the actual location and shares the perturbed location. At timestamp $t$, let $p_t^-$ and $p_t^+$ respectively represent the prior and posterior probability distributions, with $p_t^-[i]$ denoting the prior probability of location $s_i$ in the location alphabet $\mathcal{G}$, and $p_t^+[i]$ corresponding to $s_i$'s posterior probability.
To share a noisy location, PIM calculates the prior probability distribution $p_t^-$ as $p_t^- = p_{t-1}^+M$, where $M$ denotes the transition matrix. Based on the prior probabilities, it builds a $\delta$-location set $\Delta X_t$ that contains minimum number of locations with the probability sum larger or equal to $1-\delta$, i.e., $\Delta X_t=min\{s_i|\sum_{s_i}p_t^-[i]\geq 1-\delta\}$, which means a subset of locations with the total probability less than $\delta$ is omitted. After that, PIM releases the perturbed location given $\Delta X_t$ at timestamp $t$, and calls it $z_t$. The posterior probability distribution is then updated as 
$\label{eq:posterior}
    p_t^+[i]=Pr(\textbf{\textit{u}}_t^*=s_i|\textbf{\textit{z}}_t)=\frac{Pr(\textbf{\textit{z}}_t|\textbf{\textit{u}}_t^*=s_i)p_t^-[i]}{\sum_jPr(\textbf{\textit{z}}_t|\textbf{\textit{u}}_t^*=s_j)p_t^-[j]}
$
for each location $s_j$, where $u_t^*$ is the true location at timestamp $t$. 

The PIM generation can be summarized as follows:
\begin{enumerate}
  \item Generates a convex hull $K'$ from $\Delta X_t$;
  \item Builds a set $\Delta V_t$ by
$$  \Delta V_t=  \cup_{\textbf{\textit{v}}_1, \textbf{\textit{v}}_2 \in \text{ vertices of } K'} (\textbf{\textit{v}}_1 - \textbf{\textit{v}}_2)$$
  \item Forms a sensitivity hull (a convex hull) $K$ from $\Delta V_t$
, which is a stricter sensitivity metric in two dimensions than the $l_1$ norm~\cite{xiao2015protecting};
  \item Converts $K$ into isotropic position $K_I$~\cite{xiao2015protecting};
  \item Samples a point $\textbf{\textit{z}}'$ from $K_I$ using the $k$-norm mechanism~\cite{hardt2010geometry}, i.e., the probability of each point $\textbf{z}$ is
  $$Pr(\textbf{z})=\frac{1}{\Gamma(d+1)\text{VOL}(K_I/\epsilon)}exp(-\epsilon||\textbf{z}-\textbf{x}^*||_{K_I})$$, where $\textbf{x}^*$ is the true answer, $||\cdot||_{K_I}$ is the Minkowski norm of $K_I$, $d$ is the dimension ($d=2$ in the location setting), $\Gamma()$ is Gamma function and VOL$()$ is the volume, and $\epsilon$ is the privacy budget;
  \item Converts $\textbf{\textit{z}}'$ back to the original space as $\textbf{\textit{z}}$ and releases it as the final output at timestamp $t$.
\end{enumerate}

By observing the output at each timestamp and knowing the transition matrix as auxiliary information, the attacker cannot infer the actual locations since the generation process models the attacker in the exact same way.
This mechanism achieves $\epsilon$-differential privacy for the trajectories in the location datasets. For further details, we refer the reader to the original paper~\cite{xiao2015protecting}.

\vspace{-1mm}
\section{Problem Statement}
\label{sec:problems}

In this section, we describe the system setting, including the data model, the system model, and the threat model. Table ~\ref{tab:symbol} shows the commonly used notations in the paper.

\begin{table}[t]
\footnotesize
  \centering
\caption{\label{tab:symbol}Symbols and notations.}
\vspace{-3mm}
\begin{tabular}{|p{2.9cm}|p{4.8cm}| }
 \hline
 $\mathcal{X}=[x_1,x_2,\dots,x_{|\mathcal{X}|}]$ & A trajectory \\ 
 $\hat{\mathcal{X}} =[\hat{x_1},\hat{x}_2,\dots,\hat{x}_{|\hat{\mathcal{X}}|}]$ & The trajectory released by the differential privacy mechanism\\ 
 $\mathcal{X}^*=[x^*_1,x^*_2,\dots,x^*_{|\mathcal{X}^*|}]$ & The trajectory released by the post-processing  \\ 
 $\mathcal{X}'_j=[x'_{1j},x'_{2j},\dots,x'_{|\mathcal{X}'|j}]$ & The fingerprinted trajectory of the data analyzer $DA_j$  \\ 
 $\mathcal{Y}=[y_1,y_2,\dots,y_{|\mathcal{Y}|}]$ & The leaked trajectory \\ 
 $\mathcal{G}$ & The location alphabet \\ 
 $m$ & The trajectory length, i.e., $|\mathcal{X}|$  \\ 
 $p$ & The fingerprinting ratio \\ 
 $n$ & The number of data analyzers \\ 
 \hline
\end{tabular}
\vspace{-3mm}
\end{table}

\vspace{-2mm}
\subsection{Data Model}
\label{sec:data_model}

We introduce the data model for our system, including the format of trajectories, discretization, and correlations.

\vspace{-1.5mm}
\subsubsection{Trajectories}
A trajectory $\mathcal{X}=[x_1,x_2,\dots,x_{|\mathcal{X}|}]$ is an ordered sequence of location data points with the same time interval between any adjacent location points. In our setting, a location point $x$ consists of GPS coordinates only, since we pre-process the trajectories to have uniform time interval and thus omit the timestamps. Although some secondary metadata can occur such as velocities and directions, we leave these to future work.

\vspace{-1.5mm}
\subsubsection{Map Discretization}
In location settings, a map area is often discretized into cells for simplicity~\cite{gursoy2018differentially, xiao2015protecting, he2015dpt, chen2012differentially}. Following those works, we divide the continuous two dimensional space using a uniform grid of $N \times N$. Throughout the rest of the paper, we still use the term "points" to represent a cell of the grid for generalization.
\vspace{-1.5mm}
\subsubsection{Correlations}
\label{sec:correlation}
We build our correlations using the Markov chain. For each location $g \in \mathcal{G}$, the transition probability of the $k$-gram model is represented as $Pr[x_k|x_{k-1},x_{k-2},\cdots ,x_1]$. We use $2$-gram model in our scheme ($k=1$). We provide a discussion about the correlation model in Section~\ref{sec:corr_discussion}.

\vspace{-2mm}

\subsection{System Model}
\label{sec:system_model}

\begin{figure}[h]
 \centering
  \includegraphics[width=0.45\textwidth]{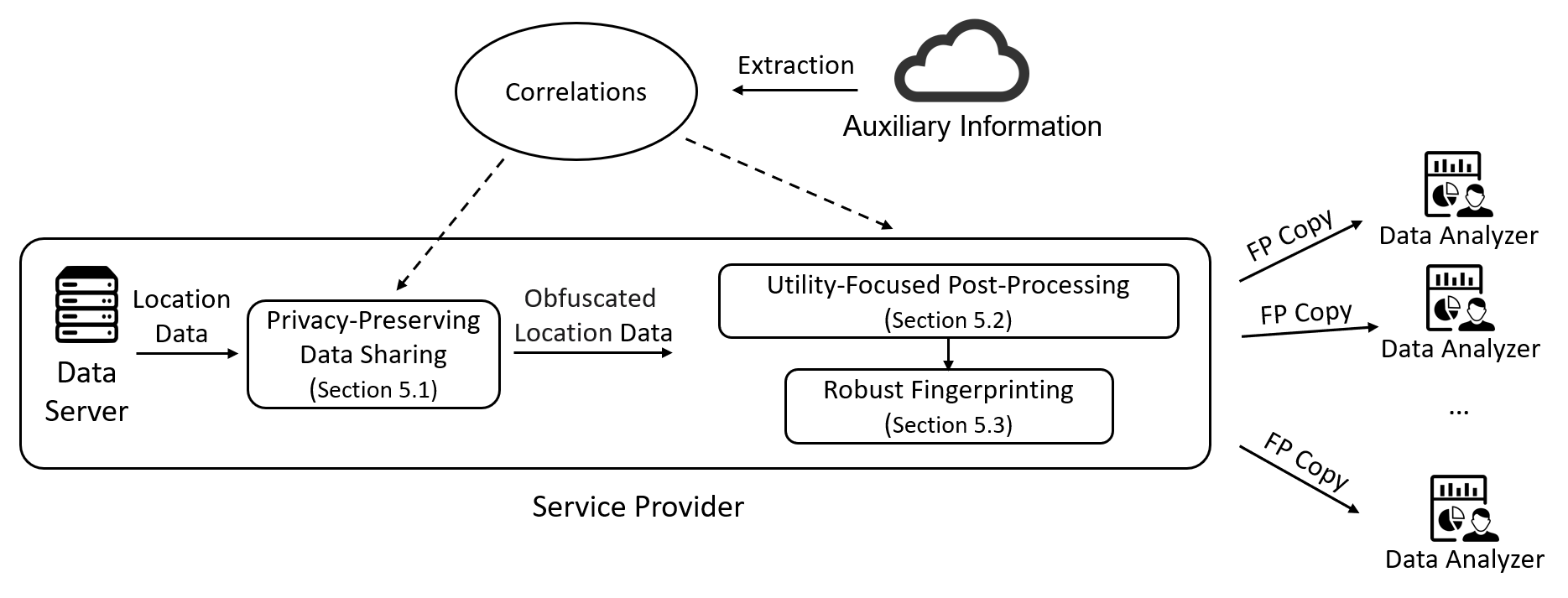}
  \vspace{-2mm}
  \caption{The system model.}
  \vspace{-1mm}
\label{fig:model}
\end{figure}

The general workflow of the framework is shown in Figure~\ref{fig:model}. There are two parties in our setting: a service provider and several data analyzers. The service provider, e.g., Google Maps or a carpooling application, collects users' location trajectories while offering the corresponding service(s) to the users. The service provider stores the location dataset in their data server and is willing to share them with other parties. Meanwhile, researchers and businesses, categorized as data analyzers, want to access such location datasets. As discussed, releasing location data may raise privacy concerns. 
Therefore, the service provider aims to ensure users' location privacy before sharing. More specifically, it can apply a privacy-preserving approach that prevents recipients (data analyzers) from knowing the users' exact locations. 
This process inevitably perturbs the data and influences data utility, which is not desired by the analyzers, especially when strong protection is applied. To best serve the analyzers and keep the users' privacy intact simultaneously, we propose a utility-focused post-processing scheme at the service provider to partially regain data utility.

% Knowing the importance of user's data privacy, they accept some degradations in usability, but they hope the data utility of the shared copy can be as high as possible.

As also discussed, a misbehaving data analyzer may distribute (leak) a copy of the received location dataset to other unauthorized parties without permission. 
Hence, we propose a novel fingerprinting scheme for location trajectories, which embeds unique fingerprint patterns into each shared location dataset. The proposed scheme is robust in case the attacker tries to distort the fingerprint by exploiting the correlations among the location data from public sources or by colluding with other misbehaving data analyzers who also receive the same location dataset (with different unique fingerprint patterns). 
Furthermore, we convert the utility-focused post-processing method into a sampling strategy and integrate it into the fingerprinting scheme. In this way, we manage to mitigate  utility degradation if differentially private mechanisms are applied in the shared dataset. 

\begin{figure}[ht]
 \centering
 \vspace{-2.5mm}
  \includegraphics[width=0.45\textwidth]{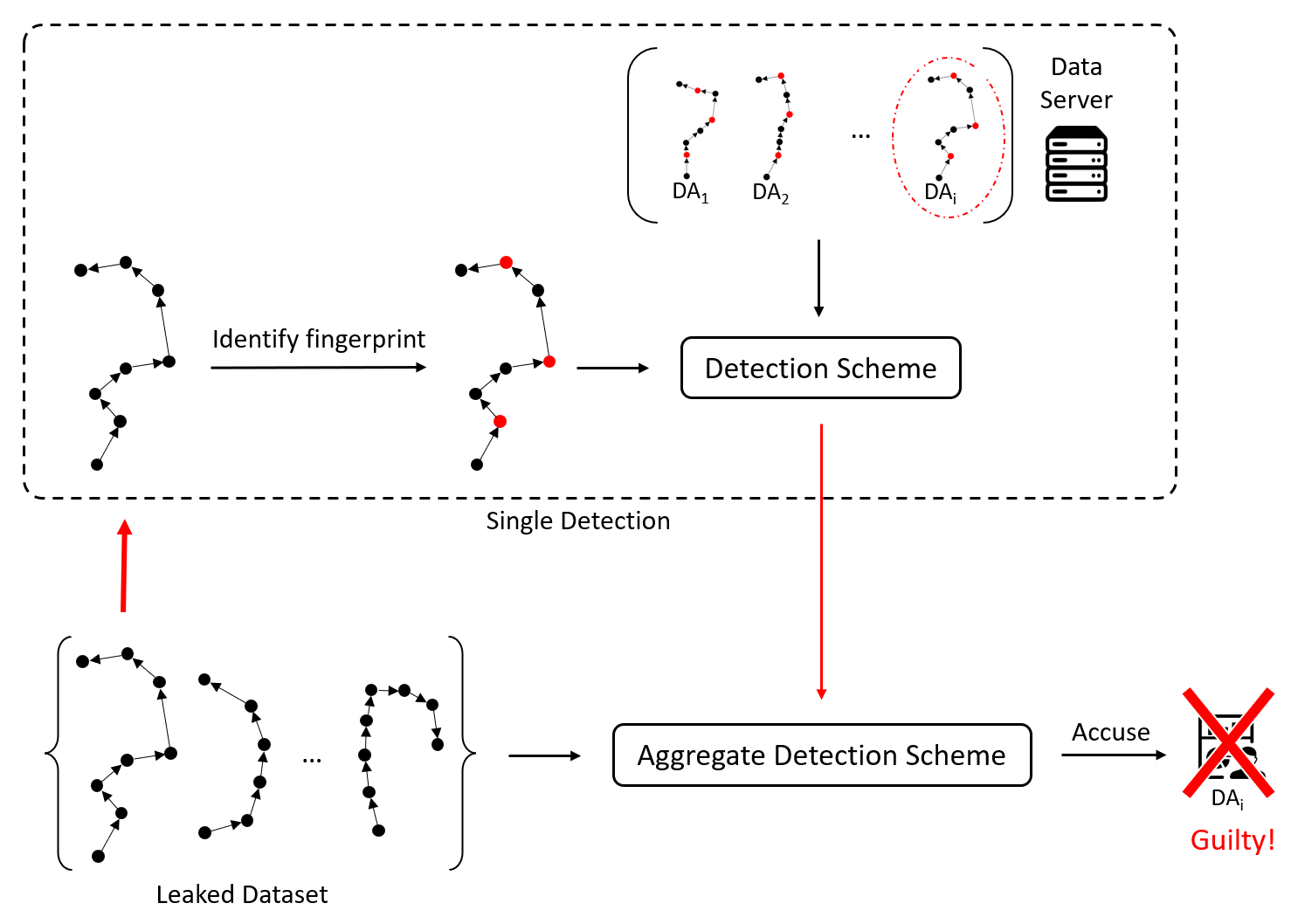}
  \vspace{-1mm}
  \caption{Detecting the source of the unauthorized redistribution.}
  %\vspace{-0.5mm}
\label{fig:detection}
\end{figure}

The fingerprint detection workflow (for the source of an unauthorized redistribution) is shown in Figure~\ref{fig:detection}. Once a location dataset is found publicly or from unauthorized sources, the service provider performs an aggregate detection scheme to identify the source of the leakage. More specifically, it runs the detection scheme for each trajectory in the leaked dataset.
The service provider aggregates the detection results (a set of accused analyzers) and finally accuses an analyzer of leaking the dataset by majority voting. The details are given in Section~\ref{sec:detection}.

\vspace{-2mm}
\subsection{Threat Model}
\label{sec:threats}
In this section, we introduce the threat model considering the parties in our system. 
The service provider is the only entity that has access to unperturbed data from the users. We assume the service provider is trusted (i.e., it does not distribute users' data to other unauthorized parties). 
The proposed scheme can be easily extended to provide privacy of users' data during the sharing process with the service provider (we discuss the practicality of a decentralized setting in Appendix~\ref{sec:decentral_discussion}).

The analyzers can be malicious. An honest analyzer never shares the fingerprinted copy that is protected under a privacy-enhancing mechanism to unauthorized parties, and it does not want to know about the original dataset. An attacker, i.e., a malicious analyzer, is curious about the original (non-perturbed) data values in the received dataset and wants to break the location privacy guarantee. For this, they can utilize auxiliary information from public sources, e.g., correlations in the map area of interest.
With the help of those information, they analyze the received trajectories and try to infer the original location points. 

On the other hand, from the perspective of fingerprinting, the attacker may want to redistribute only one trajectory or a subset of the location dataset (i.e., multiple trajectories) to other parties, e.g., motivated by profit. To avoid tracking, the attacker tries to distort the fingerprint signature. They can exploit public correlations, collude with other analyzers, or even use both to hide their identities. In the rest of the section, we discuss all the attacks the analyzers can perform against the proposed fingerprinting scheme.

\vspace{-1.5mm}
\subsubsection{Random Flipping Attack}
Random flipping attacks are the baseline attack in which the attacker distorts the location points in the trajectory in order to distort the fingerprint. For each location point in the trajectory, the attacker chooses to report another point from the actual point's neighbors with probability $p_r$. Otherwise, the attacker does not change the point and report the actual point instead.

\vspace{-1.5mm}
\subsubsection{Correlation-Based Flipping Attack}
The attacker can utilize the public correlations to improve the baseline distortion. This attack was first introduced in~\cite{yilmaz2020collusion}.
%, and we name this attack the correlated distortion attack. 
In this attack, the attacker analyzes the correlations between contiguous points along the trajectory from the start to the end. 
It checks the $2$-gram transition from the previous point to the current one, i.e., $Pr(x_1 = x_j|x_0 = x_{j-1})$ at position $j$ in the trajectory. If the transition probability is lower than a threshold $\tau$, the attacker considers the point is fingerprinted with high probability. The attacker decides to distort the point with probability $p_c$. The attacker first constructs a set containing all highly probable locations, i.e., the transition probability from the previous point $x_{j-1}$ to each point in the set is at least $\tau$. The attacker %proportionally  
samples an output based on the transition probability from the last true point $x_{j-1}$ to each point in the set. By doing so, the attacker distorts the suspicious positions, and thus avoids being detected.

If multiple parties collude by sharing their copies with each other, they can perform more powerful attacks. We consider two types of collusion attacks in our setting, differing in whether the attackers take auxiliary information into account.

\vspace{-1.5mm}
\subsubsection{Majority Collusion Attack~\cite{boneh1998collusion}}

In the majority collusion attack, the attackers collude and analyze the merged dataset point by point. At each position, the attackers always choose the most frequent value as the output. The majority voting makes the trajectory lose some fingerprint bits, which may mislead the fingerprint detection mechanism and result in accusing an innocent party.

\vspace{-1.5mm}
\subsubsection{Probabilistic Collusion Attack~\cite{yilmaz2020collusion}}

Similar to correlation-based flipping attacks, probabilistic collusion attacks~\cite{yilmaz2020collusion} exploits the auxiliary information. The attackers share the datasets and analyze them using correlations, i.e., the transition probabilities.
They also set a probability $p_e$ to approximate the actual fingerprinting probability $p$. 
Suppose the attackers are deciding the output for the $j$-th position in a trajectory.
The attackers collect all the location at position $j$ to form an alphabet $G = \{g_1, g_2, \dots, g_K\}$ at this position, where $K$ is the number of the distinct locations, and count the occurrence as $c_{j,k}$ for each location $g_k$, $k\in[1,K]$. The attackers filter those with low transition probabilities from the last released point $y_{j-1}$. Among the remaining set, they perform the probabilistic sampling, where the probability is proportional to $(1-p_e)^{c_{j,k}} \cdot (\frac{p_e}{|G_j|-1})^{n-c_{j,k}} \cdot P(x_j = g_k | x_{j-1} = y'_{j-1})$, where $G_j$ refers to the alphabet at position $j$. The first part $(1-p_e)^{c_{j,k}} \cdot (\frac{p_e}{|G_j|-1})^{n-c_{j,k}}$ is the probability of $g_k$ being the original location at position $j$ based on the assumed probability $p_e$, and the latter part is the transition probability from the previous location. By combining the two parts, the attackers are able to calibrate such probability that a location with a very low probability is barely the true location even it occurs multiple times, and a location with a high probability in the correlation model is more likely to be the true value although it occurs rarely. The attackers finally sample a location based on the weighted probability distribution and report that location at position $j$.

\vspace{-1.5mm}
\subsubsection{Re-Fingerprinting Attack}
The attacker can execute the proposed fingerprinting scheme on the fingerprinted copy in order to perturb some embedded fingerprint points, namely re-fingerprinting attack. We consider that the attacker applies the fingerprinting scheme on the received dataset using different fingerprinting ratio $p_a$.
\vspace{-0.5mm}
\section{Methodology}
\label{sec:methodology}
\vspace{-0.5mm}

We follow the following steps for each trajectory in the dataset. 
First, we protect the location datasets using a differentially private mechanism, i.e., the planar isotropic mechanism (PIM)~\cite{xiao2015protecting}. After generating the differentially private dataset, we maximize the data utility of the shared dataset by applying a post-processing strategy and further integrate it into our probabilistic fingerprinting scheme. 
In the rest of this section, we provide the technical details of these mechanisms. In Section~\ref{sec:apply_dp}, we briefly explain the reason for choosing PIM as the building block and also comparing it with other existing approaches. In Section~\ref{sec:postprocess}, we introduce the post-processing scheme that regains pairwise correlations in the differentially private dataset. In Section~\ref{sec:location_fingerprinting}, we propose our fingerprinting scheme and show how we integrate the post-processing scheme into our sampling process. In Section~\ref{sec:detection}, we show how we detect an attacker. In addition, we prove that our scheme does not violate the differential privacy guarantee provided by the differentially private mechanism in Appendix~A.

\vspace{-2mm}
\subsection{Privacy-Preserving Location Data Sharing}
\label{sec:apply_dp}
\vspace{-0.5mm}
We choose the planar isotropic mechanism~\cite{xiao2015protecting} (PIM) as the building block to ensure trajectories' privacy considering its three main advantages. First, PIM publishes trajectories with timestamps, while other approaches (e.g.,~\cite{gursoy2018utility}) do not.
By preserving timestamps, PIM is able to provide more meaningful location trajectories, enhancing their overall value.
Second, PIM and our proposed scheme share the same public information model, i.e., a correlation model generated from public sources. 
Third, as a perturbation-based method, PIM provides greater flexibility in selecting an appropriate noise level to balance privacy and utility. 
For instance, a user can either generate a noisy output with low data utility to services that have low utility requirements or release a less noisy one with high data utility to utility-sensitive services.
Synthetic methods, in contrast, only preserve statistical features and omit other essential aspects (e.g., user-specific details), which leads to a significant loss of data utility even if a high privacy budget is allocated. 

Note that we do not generate a differentially private copy for each data analyzer. In our scheme, we apply the planar isotropic mechanism (PIM) only once for each trajectory in the dataset. After that, the same noisy dataset generated from the differentially private mechanism is used throughout the entire fingerprinting process. 
This is because sharing multiple outputs on the same input under differential privacy results in cumulative privacy loss~\cite{dwork2014algorithmic}, and this may be exploited if the attackers collude and perform averaging attacks to recover the original dataset. As a result, we choose to apply PIM once for each trajectory and then use the same noisy copy in our proposed fingerprinting scheme.

Similar to other perturbation-based approaches that ensure event-level differential privacy, PIM generates high amount of noise for each location point under high privacy protection, leading to significant utility loss in the the shared location dataset. Since there are no solid constraints for the neighboring locations in the released trajectory that guarantee the moving patterns are realistic, the pairwise correlations inside are mostly very low for common $\epsilon$ values. Influenced by the two aforementioned factors, data utility of the whole trajectory decreases significantly. In other words, the trajectories before and after perturbations differ considerably in terms of shape and point-wise relations. As a result, the dataset is almost unusable for the data analyzers as they can hardly infer meaningful pieces of information, e.g., moving trends and statistics, from the trajectories.
To solve this problem, we propose our post-processing scheme, called utility-focused post-processing.

\vspace{-2mm}
\subsection{Utility-Focused Post-Processing}
\label{sec:postprocess}
\vspace{-0.5mm}
 %Thus, here, our goal is to generate a trajectory from the noisy release with better data utility and use it as the true value in fingerprinting.

Utility-focused post-processing scheme utilizes the auxiliary information that is also used in PIM and from public sources to boost data utility of the released trajectory data. We start with the definition of the $\tau$-probable set in Definition~\ref{def:tau_prob_set}. 

\begin{definition}[$\tau$-Probable Set]
\label{def:tau_prob_set}
 Let $\tau \in [0,1]$ and $\mathcal{G}$ be the set of discrete map areas. $\mathcal{M}$ is the $2$-gram Markov model. Given a location point $g^* \in \mathcal{G}$, the $\tau$-probable set of $g^*$ is defined as 
 \begin{equation}
 \label{eq:prob_set}
     prob_\tau(g^*) \leftarrow \{g | Pr[x_1 = g|x_0 = g^*] \geq \tau\}, g \in \mathcal{G}
 \end{equation}
 , where $Pr[x_1 = g|x_0 = g^*]$ is the transition probability obtained from the correlation model $\mathcal{M}$. 
\vspace{-1mm}
\end{definition}

The idea of $\tau$-probable set origins from~\cite{yilmaz2020collusion}, where the authors only consider pairwise data points with transition probability larger or equal to $\tau$. We build the correlations using the $2$-gram Markov chain (following~\cite{yilmaz2020collusion}) and consider the transitions based on the previous locations in the trajectory.
\vspace{-0.5mm}

In the post-processing scheme, we iterate the location points in the trajectory in a sequential order. While post-processing the $j$-th location point of a differentially private trajectory, named $\hat{x}_j$, we first obtain the $(j-1)$-th output $x^*_{j-1}$ that is generated from the post-processing scheme and calculate its $\tau$-probable set $prob_\tau(x^*_{j-1})$. If $\hat{x}_j$ is in $prob_\tau(x^*_{j-1})$, the correlations are preserved between the two points, and thus we do not modify the points. Otherwise, the correlations do not exist. In this case, we choose the one that is the closest to $\hat{x}_j$ within the $\tau$-probable set as the output. The new point $x^*_j$ is treated as the original value of the corresponding data point during the fingerprinting process.

\begin{figure}[h]
 \centering
  \includegraphics[width=0.23\textwidth]{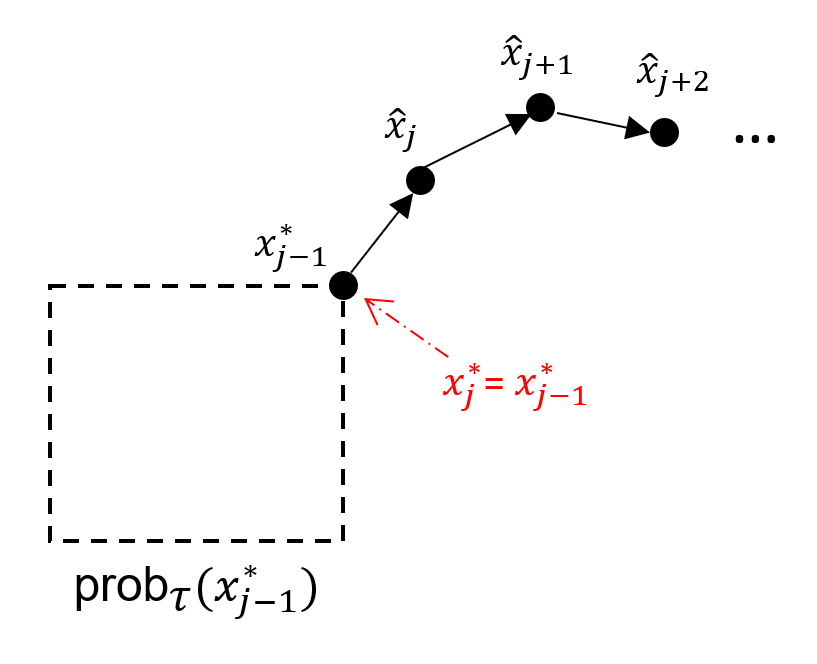}
  \vspace{-2mm}
  \caption{\textit{Pit falling}. $x_{j-1}^*$ is the last smoothed point. The following outputs $x_j^*, x_{j+1^*},\cdots$ will stay at the same position as $x_{j-1}^*$, forming a pit.}
  \vspace{-5mm}
\label{fig:pit}
\end{figure}

Notice that selecting the points in the $\tau$-probable set depends on the transition probability from the correlation model. Thus, it is not guaranteed that the $\tau$-probable set is a circle-like shape covering all the directions of the previous location $\hat{x}$. Due to the insufficiency of the correlations generated from the publicly available datasets, in some extreme cases, there exist no suitable location points in the set getting close to $\hat{x}_j$ compared with the previous location $x^*_{j-1}$. If this happens and the trajectory trend continues, i.e., no turning back, the following outputs will fall into a pit. 
Figure~\ref{fig:pit} is an example of pit falling. $x^*_{j-1}$ is the post-processed output at position $j-1$, and the $\tau$-probable set of it is marked using a dashed square. While deciding $x^*_j$, the scheme finds that $\hat{x}_j$ is outside of the $\tau$-probable set, and thus it should choose the closest point to report. 
As the closest location is identical to the previous release $x^*_{j-1}$, the algorithm still reports $x^*_j = x^*_{j-1}$. $x^*_j$ and the remaining points $x^*_{j+1},x^*_{j+2},\cdots$ stays at the same position following the same process, making the trajectory fall into a pit. 
Our solution is to let $x^*_j = \hat{x}_j$ in this case. By doing so, we force the scheme to jump out of the pit while the generation still follows the temporary trend of the trajectory. The complete algorithm of the post-processing scheme is shown in Algorithm~\ref{alg:smoothing} in Appendix \ref{app:postprocess}.

\vspace{-1.5mm}
\subsection{Robust Fingerprinting}
\label{sec:location_fingerprinting}
\vspace{-0.5mm}
Traditional fingerprinting approaches~\cite{boneh1998collusion, tardos2008optimal} do not consider spatial/temporal correlations and they treat each point independently. However, location points in a trajectory are highly correlated, especially the neighboring ones. Thus, modifying a location point without following the correlation model will make a point far away from its neighboring points such that the attacker can easily identify most of the fingerprint bits by checking pairwise correlations. The probabilistic fingerprinting scheme (PFS)~\cite{yilmaz2020collusion} is the only existing approach that takes correlations into account during fingerprinting. However, the scheme in~\cite{yilmaz2020collusion} requires the states of the data to be limited and intertransitable. If the number of states are large and they have sparse correlations, i.e., transitions only exists between a small portion of the state pairs, ~\cite{yilmaz2020collusion} starts having limitations. In addition, PFS does not consider the privacy of the shared data streams. In the following, we first briefly introduce PFS.

\vspace{-1mm}
\subsubsection{The Probabilistic Fingerprinting Scheme (PFS)}
\label{sec:pfs}
PFS embeds the fingerprint codes from the start to the end of a data stream, i.e., $x_0$ to $x_{|\mathcal{X}|-1}$. Suppose we are generating the $j$-th position in a data stream $\mathcal{X}$, and the fingerprinting ratio is $p$. 
While determining the output $x'_j$, PFS checks the transition probability $Pr[x_j = g|x_{j-1} = x'_{j-1}]$ for each $g$ in the alphabet and filters those with low probability (i.e., lower than a threshold $\sigma$).
PFS then forms a probability distribution  among the remaining values. If the original value is not eliminated, $Pr[x_j]$ is set to $1 - p$ with the remaining $p$ is proportionally assigned to the rest according to their transition probabilities. If the original value of the corresponding data point at position $j$ is eliminated, the scheme only generates the output proportionally from the remaining values.

\begin{figure}[h]
 \centering
  \includegraphics[width=0.28\textwidth]{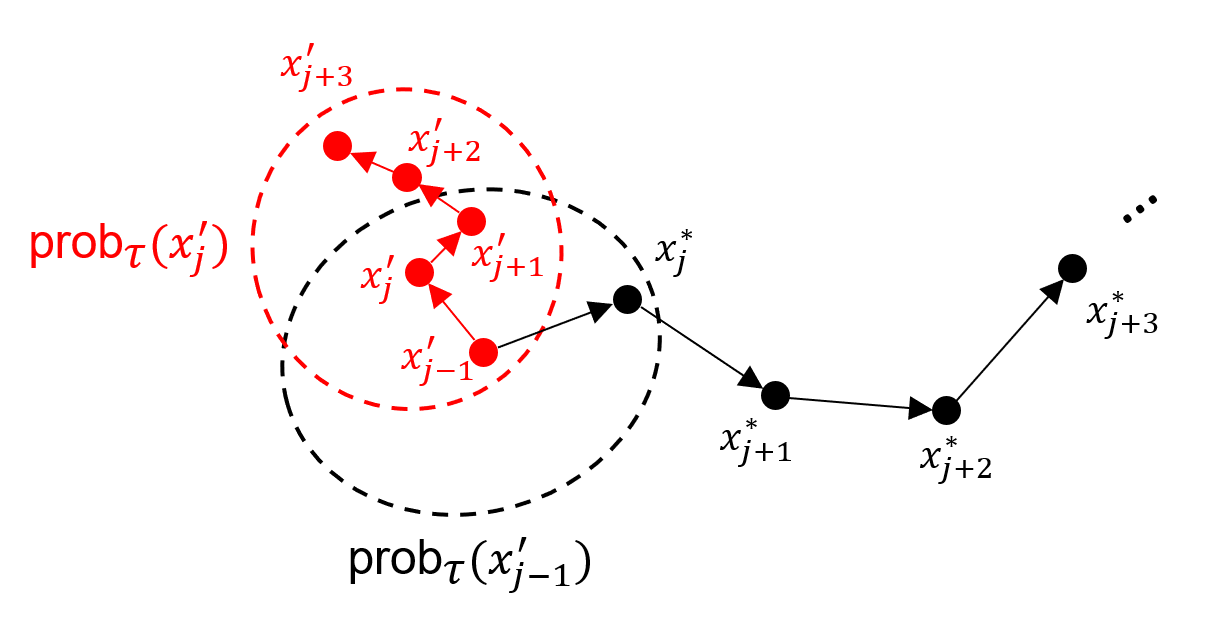}
  \caption{\textit{Forced deviation}. $x_j'$ is sampled inside the $\tau$-probable set of $x_{j-1}'$, but it makes the next original value of the corresponding data point $x_{j+1}^*$ outside its $\tau$-probable set. Following FPS, the following points will be sampled among $prob_\tau(x_j')$ only.}
\label{fig:deathloop}
\end{figure}

\begin{figure}[h]
         \centering
         \includegraphics[width=0.45\textwidth]{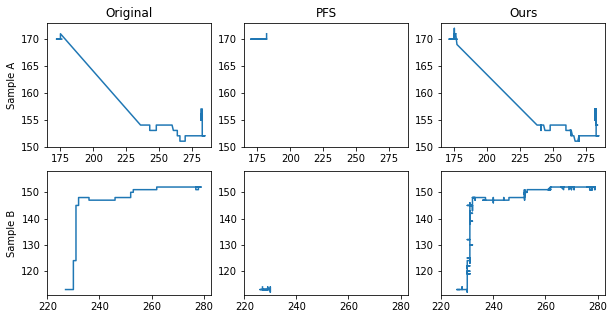}
         \caption{Visualization of two fingerprinting schemes, i.e., (i) PFS~\cite{yilmaz2020collusion} and (ii) our scheme, on two trajectory samples. Forced deviation is clearly shown in the generated copies using PFS.}
         \label{fig:forced_deviation_sample}
         \vspace{-5mm}
 \end{figure}

However, PFS cannot be applied to location datasets even without privacy protection. The most critical problem is forced deviation.
PFS process normally works in location fingerprinting, but when the correlations are low between the data points, it starts to show limitations. According to PFS, the scheme eliminates the original value of the corresponding data point if the correlations do not hold. Then, the scheme proportionally samples a point from the remaining $\tau$-probable set consisting of highly probable points and reports that one.
In trajectory fingerprinting, once the sampled output appears outside of the next point's $\tau$-probable set, the rest of the points will wander around the $\tau$-probable set forever. We show this in Figure~\ref{fig:deathloop} as an example. 
Here, PFS fingerprints the $j$-th position in the trajectory, while $x'_{j-1}$ is the last fingerprinted location and the dashed circular area in black is the $\tau$-probable set of $x'_{j-1}$. $x^*_j$ is the actual location at position $j$, and it is in the $\tau$-probable set of $x'_{j-1}$. PFS wants to sample a point among the $\tau$-probable set and releases that point. If the sampled point is located as $x'_j$ in Figure~\ref{fig:deathloop}, we realize that the next original value of the corresponding data point $x^*_{j+1}$ is not in the $\tau$-probable set of $x'_j$. 
In this case, the scheme will sample a location only among the set, regardless of the distance from the original value.
The next original value $x^*_{j+2}$ will be more likely to occur outside of the $\tau$-probable set (marked by red dashed circle) as well since the actual trajectory moves forward and the sampled output sticks to the area close to the first separation, i.e., $x'_j$. If the generation continues, the fingerprinted locations will be sampled around the first deviated location $x'_j$, and this will finally result in forced deviation.
We show some examples for this scenario for better clarification by applying PFS and our proposed scheme on two trajectory samples in Figure~\ref{fig:forced_deviation_sample}. As shown, PFS falls into forced deviation at the very beginning for each sample, while our approach generates fingerprints along the trajectory (i.e., the right figures).

\subsubsection{Direction-Sensitive Fingerprinting Scheme For Location Trajectories}
\label{sec:DSFS}
To solve the aforementioned challenges, we propose a new sampling scheme, called the direction-sensitive fingerprinting scheme (see Algorithm~\ref{alg:dsfs} for details). For a released point $x_{j-1}'$, we first form a set containing all locations closer or equal to $x^*_j$ than $x_{j-1}'$ in the $\tau$-probable set, called \textbf{$\tau$-closer set}, which can be expressed as 
$$prob_\tau^c(x'_{j-1}) \leftarrow \{g \big\vert \lVert  g, x^*_j \rVert_2 \leq \lVert  x'_{j-1} , x^*_j\rVert_2, g \in prob_\tau(x'_{j-1})\}$$. Normally, if the original value of the corresponding data point $x^*_j$ is in the $\tau$-closer set, we sample the output among it by setting the probability of choosing the original value as $1-p$ and the rest is proportionally assigned based on the transition probability to the destination. We improve the sampling process to avoid forced deviation during the generation. There are four cases while selecting the original value at the $j$-th position. If the original value $x^*_j$ is in the $\tau$-closer set of the previously released location $x'_{j-1}$, there is no difference between ours and in PFS. If $x^*_j$ is not in the $\tau$-closer set, we check its membership in the $\tau$-probable set and sample from the same distribution as above, but among the $\tau$-probable set instead. If not, we check the closest point $\tilde{x}$ to the original value $x^*_j$ in $prob_\tau(x'_{j-1})$. If $\tilde{x}$ is the same as $x'_{j-1}$, which means there exists no such location closer than $x^*_j$, we let the temporary true value be $x^*_j$. Otherwise, we choose $\tilde{x}$ as temporary original value at this timestamp and perform the proportional sampling scheme. 
For the first location $x^*_0$ in the trajectory, we do not have conditional probabilities. Instead, we use the emission probability of $x^*_0$'s neighboring locations, i.e., $Pr(g) = \frac{(\text{\# of points at }g)}{\sum_{g'} (\text{\# of points at }g'), g'\in neigh(x^*_0)},  g \in neigh(x^*_0)$, where $neigh(x)$ denotes a set of all neighbors of $x^*_0$ (including $x^*_0$ itself), in the sampling process.% alternatively.

In order to offer fingerprint robustness and data usability at the same time, we integrate the proposed post-processing scheme in Section~\ref{sec:postprocess} into our fingerprinting. In particular, if the next original value $x^*_j$ is not in the $\tau$-probable set, we follow the post-processing scheme to choose the closest point as the surrogate, and assume it to be the original value. This post-processing integration does not take effect if we work on not differentially private trajectories, as pairwise correlations are preserved along those trajectories. If dealing with noisy trajectories, i.e., protected under differential privacy, the post-processing step will regain pairwise correlations and thus improve data utility for location datasets. 

\begin{algorithm}[h] 
\small

 \SetKwInOut{Input}{input}
 \SetKwInOut{Output}{output}
  \Input{Trajectory $\mathcal{X^*}=[x^*_1,x^*_2,\dots,x^*_m]$, location alphabet $\mathcal{G}$, emission probability $Pr[g_p]$ and transition probability $Pr[g_q|g_r]$ for any locations $g_p, g_q, g_r \in \mathcal{G}$, probability threshold $\tau$, fingerprinting ratio $p$, ratio balancing factor $\theta$, the first fingerprinted trajectory $\mathcal{X^o}=[x^o_1,x^o_2,\dots,x^o_m]$}
 \Output{Fingerprinted trajectory $\mathcal{X}'=[x'_1,x'_2,\dots,x'_m]$}
    % \uIf{$1 \in BS$}{
    %     \uIf{$F_1 == true$}{
    %         $x'_1 \leftarrow x^o_1$ \;
    %     }
    %     \uElse{
    %         $x'_1 \leftarrow x^*_1$ \;
    %     }
    % }
    % \uElse{
        $PD \leftarrow$ $Pr[x'_1 = x^*_1] = 1 - p_{current}, Pr[x'_1 = g] = \frac{Pr[g]}{\sum_{g'\in \mathcal{G} \backslash x^*_1 } Pr[g']},  g\in \mathcal{G}$\;
        $x'_1 \leftarrow$ sample from $PD$\;
    % }
    $p_{current} = p$\;
    \ForAll{$j \in 2, 3, \dots, m$}{
        % \uIf{$j \in BSI$}{
        %     \uIf{$F_j == true$}{
        %         $x'_j \leftarrow x^o_j$ \;
        %     }
        %     \uElse{
        %         $x'_j \leftarrow x^*_j$ \;
        %     }
        %     continue \;
        % }
        $prob_\tau(x'_{j-1}) \leftarrow \tau$-probable set of $x'_{j-1}$\;
        $prob_\tau^c(x'_{j-1}) \leftarrow \{g| \lVert  g, x^*_j \rVert_2 \leq \lVert  x'_{j-1} , x^*_j\rVert_2, g \in prob_\tau(x'_{j-1})\}$\;
        \uIf{$x^*_j \in prob^c_\tau(x'_{j-1})$ and $|prob_\tau^c(x'_{j-1})| > 1$}{
            $PD \leftarrow$  $Pr[x'_j = x^*_j] = 1 - p, Pr[x'_j = g] = \frac{Pr[x'_j = g | x'_{j-1}]}{\sum_{g'\in prob_\tau(x'_{j-1}) \backslash x^*_j} Pr[x'_j = g' | x'_{j-1}]} * p , g \in prob_\tau^c(x'_{j-1})$\;
            $x'_j \leftarrow$ sample from $PD$\;
        }
        \uElseIf{$x^*_j \in prob_\tau(x'_{j-1})$ and $|prob_\tau^c(x'_{j-1})| == 1$}{
            $PD \leftarrow$  $Pr[x'_j = x^*_j] = 1 - p, Pr[x'_j = g] = \frac{Pr[x'_j = g | x'_{j-1}]}{\sum_{g'\in prob_\tau(x'_{j-1}) \backslash x^*_j} Pr[x'_j = g' | x'_{j-1}]} * p , g \in prob_\tau(x'_{j-1})$\;
            $x'_j \leftarrow$ sample from $PD$\;}
        \uElse{
            $x_{closest} \leftarrow $ closet point to         $x^*_j$ in $ prob_\tau(x'_{j-1})$\;
            \uIf{$|prob_\tau(x'_{j-1})| <= 1$}{
                $x'_j \leftarrow x^*_j$\;
            }
            \uElse{
                $PD \leftarrow$  $Pr[x'_j = x_{closest}] = 1 - p, Pr[x'_j = g] = \frac{Pr[x'_j = g | x'_{j-1}]}{\sum_{g'\in prob_\tau(x'_{j-1}) \backslash x_{closest}} Pr[x'_j = g' | x'_{j-1}]} * p, g \in prob_\tau(x'_{j-1})$\;
                $x'_j \leftarrow$ sample from $PD$\;
            }
        }
        
        \uIf{$j$ mod $\lceil \frac{1}{p} \rceil == 0$}{
            $count \leftarrow$ \# of fingerprinted positions\;
            \uIf{$count > p * j$}{
                $p_{current} \leftarrow p * (1-\theta)$\;
            }\uElseIf{$count < p * j$}{
                $p_{current} \leftarrow p * (1+\theta)$\;
            }\uElse{
                $p_{current} \leftarrow p$\;
            }
        }
    }

 \caption{Direction-Sensitive Fingerprinting Scheme}
 
 \label{alg:dsfs}
\end{algorithm}

In addition, we follow~\cite{yilmaz2020collusion} and use the balancing strategy. During the fingerprinting generation, some positions are perturbed while some remain the same as the original values. We use $FP$s and $NoFP$s to represent them, respectively. PFS balances the distribution of the $FP$s by using the balancing factor $\theta$. The scheme checks the $FP$ count every $\lceil \frac{1}{p} \rceil$ points. If the actual $FP$ count is larger than expected, then the temporary fingerprinting ratio is changed to $p * (1 - \theta)$. If the $FP$ are not enough, the ratio becomes $p * (1 + \theta)$. The complete algorithm is shown in Algorithm~\ref{alg:dsfs}.

\vspace{-2mm}
\subsection{Detecting the Source of the Unauthorized Redistribution}
\label{sec:detection}
\vspace{-0.5mm}

We use similarity-based detection~\cite{yilmaz2020collusion} with our improvement. During traditional similarity-based detection, data points in the leaked data are compared with the distributed copies. At each position, if the leaked data point matches some data analyzers, each of them will be assigned a score $\frac{1}{|X|}$, where $|X|$ is the length of the data. After all data points are inspected, the analyzer with the largest cumulative score is considered malicious. In location data, slight perturbation is enough to invalidate those exact matches, and thus influence the detecting accuracy. Thus, we replace it with a distance-based match in similarity-based detection. For each location point in the trajectory, we assign $\frac{1}{|X|}$ to all the points that have the shortest distance to the leaked location point instead exact matches, which significantly improves our detection.

The described detection works on a single trajectory leakage from a shared location dataset. For a multi-trajectory leakage, we implement an aggregate detection scheme to identify the source of the unauthorized redistribution. We first use the distance-based detection to analyze leaked trajectories one by one in the leaked dataset and accuse one to be malicious for each leaked trajectory. Among all the accused data analyzers, we do majority voting on them and choose the most frequent one as the final malicious data analyzers. The evaluation results of multi-trajectory leakage are in Appendix~\ref{eval:multi_leakage}.

\vspace{-2mm}
\section{Evaluation}
\label{sec:evaluation}
\vspace{-0.5mm}
We implemented the proposed fingerprinting scheme and provide the experimental results. 
We first evaluate fingerprint robustness against multiple attacks while we apply fingerprinting schemes on original datasets, which is critical as service providers do not always use differentially private mechanisms due to data utility concerns. After that, we evaluate fingerprint robustness on datasets that are protected under differential privacy. We also show that, using other differentially private mechanisms (e.g., AdaTrace~\cite{gursoy2018utility}), we still achieve similar performance against the considered attacks. In terms of data utility, we evaluate fingerprinted datasets using five utility metrics mentioned in Section~\ref{sec:utility_metrics}. Furthermore, we performed parametric experiments on trajectory length (in Appendix~\ref{eval:length}) and time complexity (in Section~\ref{eval:time}). For the experiments, we used a rack server with 64GB Memory (DDR4, 2666Mhz) and Intel Xeon E5-2650 @ 2.20GHz with $40$ cores. We ran all experiments for more than $1,000$ times with $20$ dataset shuffles and took the average.

% \subsubsection{Baseline Approaches}
% For comparison, we implement a randomized fingerprinting scheme (RFS) as the baseline approach. RFS embeds a fingerprint value at each position with a uniform probability $p$. At those selected positions, RFS generates a multivariate Laplacian noise to the original coordinates. The scales along the longitude and latitude are the same and aligned with our proposed scheme by the average deviation between corresponding points. We also implement the vanilla Boneh-Shaw codes and the vanilla Tardos codes to compare their performance with our scheme. 

\vspace{-1.5mm}
\subsection{Datasets}
We used 4 datasets during evaluation: 1) the GeoLife dataset (Version $1.3$)~\cite{zheng2010geolife}, 2) the Taxi dataset~\cite{moreira2013predicting}, 3) the Oldenburg dataset~\cite{brinkhoff2002framework}, and 4) the San Joaquin dataset~\cite{brinkhoff2002framework}. The GeoLife and Taxi datasets are real-life ones and the Oldenburg and San Joaquin datasets are synthetic ones from the Brinkhoff generator.  GeoLife dataset  contains $17,621$ trajectories generated by $182$ users using different GPS devices over five years (April 2007-August 2012), including $1,292,951$ kilometers in the distance and $50,176$ hours in time, where most of the locations are in Beijing, China. The Taxi dataset is used in Taxi Service Prediction
Challenge at ECML-PKDD 2015~\cite{moreira2013predicting}, including $1,710,670$ taxi trajectories in Porto, Portugal. The remaining two datasets are synthesized from the Brinkhoff generator for moving objects~\cite{brinkhoff2002framework} in the city of Oldenburg and San Joaquin, respectively. We generate $5,000$ trajectories for each dataset. 

\vspace{-1.5mm}
\subsubsection{Data Pre-Processing}
We pre-processed the trajectories to avoid various data intervals. We smoothed the trajectories to have similar time intervals, i.e., around $60$ seconds. For each dataset, we defined an area of interest that covers most of the trajectories and cut and filtered out the trajectory fragments outside the area. We picked $1,000$ trajectories as our fingerprinting targets and used the remaining ones to build public correlations. 
% To build a reasonable correlation model for our system, we pre-process the trajectories to limit the time interval uniformly that approximates to $3$ seconds to simulate the services such as Google Maps and carpooling.

% Moreover, we remove the altitude and construct a $2$-dimensional geographical area and define the area of interest by the latitude $\in [39.6797, 40.1280]$ and the longitude $\in[116.0287, 116.7064]$. Since there are some location points far from Beijing, 
% we remove those fragments from the dataset. We discretize the area into cells uniformly using a $1000 \times 1000$ grid both in the PIM and in the HPFS. The different selections of the grid size do not affect the performance since the two processes (PIM and HPFS) are independent, but we match the two grids for simplicity. 

% We randomly select $80\%$ of the trajectories in the dataset to form the correlation model, while the rest are used for the evaluation on fingerprint robustness and data utility. 

\vspace{-2mm}
\subsection{Experimental Settings}
\label{ex:results}
\vspace{-0.5mm}

We compare our fingerprinting scheme with two traditional ones, i.e., the Boneh-Shaw codes and the Tardos codes. We evaluate detection accuracy of the three schemes on both non differentially private and differentially private datasets. The Boneh-Shaw codes and the Tardos codes do not support detection of multiple trajectories, so we use the same detection logic as ours, i.e., working on trajectories one by one and then majority voting, to fit our experiments.

The following experiments assume that the attacker(s) will only leak one trajectory from the entire dataset. As we mentioned in Section~\ref{sec:detection}, we perform detection one by one on each leaked trajectory and do majority voting for final accusation. The detection processes of leaked trajectories are independent from each other, which makes the problem  become a combination problem (i.e., given detection accuracy for a single trajectory equal to $p$, what is the detection accuracy of $k$ trajectories using majority voting?). As we will show in the next sections, our approach significantly outperforms the existing schemes and keeps over $90\%$ detection accuracy in most cases. If multiple trajectories are leaked, the overall detection accuracy goes up and reaches to $99.99\%$. We show this in Appendix~\ref{eval:multi_leakage}. For simplicity, we only consider the leakage of one trajectory in the following.

\vspace{-1.5mm}
\subsubsection{Parameter Settings}
If not specified, we use the following parameter setting throughout the experiments. An original dataset contains $100$ randomly selected trajectories, and each has $100$ locations. We assume $100$ SPs get the copies by default. We set $\tau = 0.005$ as the correlation threshold concluded from our experiments and the fingerprint balancing factor $\theta = 0.5$. 
% For any Laplaican noise, the scale is set to $0.2$ (in terms of cell ID other than the GPS coordinates). 
The Tardos codes use $\omega = 0.01$ as the error probability. The Boneh-Shaw codeword consists of $|X|$ blocks and $1$ location points in each block. For PIM, we follow~\cite{xiao2015protecting} and set $\delta = 0.01$ for the $\delta$-location set. The fingerprinting ratio is set to $0.4$. We suppose the attacker(s) use $p_c = 0.8$ and $p_r = 0.8$ in random and correlation-based flipping attacks, respectively, and $3$ service providers collude by default.

\subsection{Evaluation Metrics}

\vspace{-1.5mm}
\subsubsection{Fingerprint Robustness Metric}
We define a successful accusation as correctly identifying the attacker who leaks the data. Our evaluation metric of fingerprint robustness is then represented as $Accuracy = \frac{\text{(\# of successful accusation)}}{\text{(\# of trials)}}$. If multiple attackers collude, we consider catching one of the colluding attackers. Since the Tardos codes focus on catching all who leak the data, we adjust the accusation process for alignment. More specifically, we only consider the one with the highest scores in the Tardos detection instead of using the threshold $20ck$ (in Appendix~\ref{sec:tardos}).

\vspace{-1.5mm}
\subsubsection{Utility Metrics}
\label{sec:utility_metrics}
Following the existing works~\cite{xiao2015protecting, he2015dpt, gursoy2018utility}, we introduce our utility metrics as follows. 

% \vspace{-1mm}
% \paragraph{Correlation Fit Rate}
% It shows how consecutive location pairs conform the correlations, i.e., $Pr(x_{j+1}|x_j) \geq \tau$ for a location point $x_j$. If the rate is high, most pairs follow the correlations; otherwise, the public correlations do not hold in the trajectories.

\vspace{-1mm}
\paragraph{Query Answering of Location Points}
\label{sec:qa_t}
The count query is one of the most frequent usages for location datasets. 
% By analyzing the error of the query outputs between the original dataset and the fingerprinted dataset, we are able to quantify the utility loss. 
Let $Q_t(D, g)$ denote the query ``how many trajectories pass a circular area represented by a center $c$ and a radius $r$ in the dataset $D$''. Then, we define the relative error as

$$AvRE =\frac{|Q_t(D,g)-Q_t(D', g)|}{max(Q_t(D,g),b)}$$
, where $D$ is the original dataset and $D'$ is the output of our scheme. We set $b=0.01 \times |D|$ according to~\cite{nergiz2008towards, zhang2016privtree, chen2012differentially, gursoy2018utility}. 
% from $4000$ random queries to evaluate the performance. We next introduce two different queries.

\vspace{-1mm}
\paragraph{Query Answering of Patterns}
We also implement another query answering metric for patterns. As discussed in Section~\ref{sec:correlation}, we only focus on the $2$-gram patterns. Given a $2$-gram pattern $P$, the count query on $P$ is $Q_p(P,D)$ that counts $P$ in the dataset $D$. We also evaluate the utility using relative error.

\vspace{-1mm}
\paragraph{Area Popularity}
We follow~\cite{gursoy2018utility} and evaluate the divergence of area popularity rankings. Based on the number of location points within each area, we generate the popularity ranking for each fingerprinting scheme. We compare the ranking with the one from the original dataset and calculate the Kendall-tau coefficient, which is defined as $KT = \frac{(\text{\# of concordant pairs}) - (\text{\# of discordant pairs})}{(\text{\# of pairs})}$. The kendall-tau coefficient measures ordinal association 
between sequences. Higher coefficient represents better utility.

\vspace{-1mm}
\paragraph{Trip Error}
Trip error~\cite{gursoy2018utility} measures trip length. We calculate the lengths of all trajectories in the dataset and put them into $11$ bins, i.e., $[0, \frac{L}{10}), [\frac{L}{10}, \frac{2L}{10}), \cdots, [\frac{9L}{10}, L)$, and $[L,\infty)$, where $L$ is the maximum trip length in the original dataset. We calculate the Jensen-Shannon Divergence (JSD) between the fingerprinted dataset and the original dataset.

\vspace{-1mm}
\paragraph{Diameter Error}
Diameter error~\cite{gursoy2018utility} is similar to trip error, but it considers distances between contiguous location points along trajectories. We use $11$ bins and then evaluate the Jensen-Shannon Divergence.

\vspace{-1mm}
\paragraph{Trajectory Similarity}
In the services like carpooling, the trajectory shape is an important feature that can be used for the service to design an optimal strategy. We use 2-dimensional dynamic time wrapping (DTW)~\cite{shokoohi2017generalizing} to evaluate the similarity between the original and the fingerprinted datasets.

% We first show the fingerprinted dataset preserves high correlation fit rate compared with the baseline method. Then, we evaluate the fingerprint robustness in the worst case, i.e., leaking only one trajectory, since the fingerprint code length may be insufficient for a successful accusation. In Section~\ref{sec:eval_dataset}, we provide the results when the attacker leaks the whole dataset instead of a single trajectory. Moreover, we show the data utility of the fingerprinted dataset. We also evaluate the time efficiency of our scheme. In Appendix~\ref{sec:exp_app}, we display additional results when fingerprinting is applied directly on the original dataset if the privacy is not considered. 

\begin{figure*}[h]
     \centering
    \includegraphics[width=\linewidth]{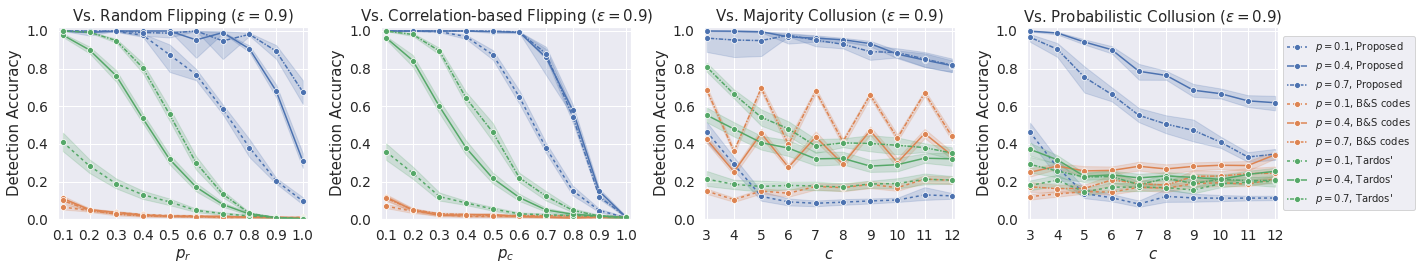}{\hspace{5.7cm}}\\
    \vspace{-0.5cm}
    \hspace{1.15cm}
    \subfloat[vs. RF]{\label{fig:raw_accuracy_a}\hspace{2cm}}
    \hfill
    \subfloat[vs. CF]{\label{fig:raw_accuracy_b}\hspace{2cm}}
    \hfill
    \subfloat[vs. MJR]{\label{fig:raw_accuracy_c}\hspace{2cm}}
    \hfill
    \subfloat[vs. PROB]{\label{fig:raw_accuracy_d}\hspace{2cm}}
    \hspace{2.3cm}
     \vspace{-3mm}
          \captionsetup{justification=centering}
          \caption{Fingerprint robustness on the non-differentially private dataset (the GeoLife dataset~\cite{zheng2010geolife}) against a) random flipping attacks, b) correlation-based flipping attacks, c) majority collusion attacks, and d) probabilistic collusion attacks using three methods, i.e., 1) our scheme (represented as "Ours" in the legends), 2) the Boneh-Shaw codes~\cite{boneh1998collusion} ("BS"), and 3) the Tardos codes~\cite{tardos2008optimal} ("TD") for different fingerprinting ratio $p$.}
          
          %with regard to accuracy and utility, i.e., (a) against random distortion attacks on the noisy dataset, (b) against random distortion attacks for different privacy levels $\epsilon$, and (c) against correlated distortion attacks for different privacy levels $\epsilon$.}}
        \label{fig:raw_accuracy}
        \vspace{-2mm}
\end{figure*}

\begin{figure*}[h]

    \centering
    \includegraphics[width=\linewidth]{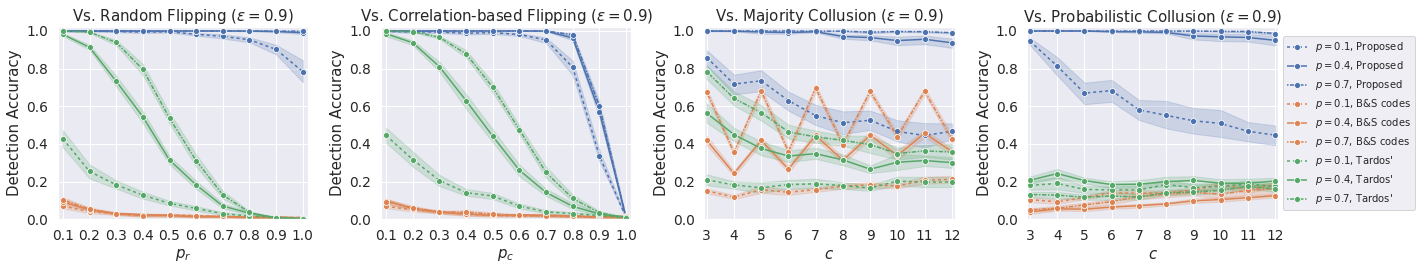}{\hspace{5.7cm}}\\
    \vspace{-0.5cm}
    \hspace{1.15cm}
    \subfloat[vs. RF]{\label{fig:dp_accuracy_a}\hspace{2cm}}
    \hfill
    \subfloat[vs. CF]{\label{fig:dp_accuracy_b}\hspace{2cm}}
    \hfill
    \subfloat[vs. MJR]{\label{fig:dp_accuracy_c}\hspace{2cm}}
    \hfill
    \subfloat[vs. PROB]{\label{fig:dp_accuracy_d}\hspace{2cm}}
    \hspace{2.3cm}
     \vspace{-3mm}
          \captionsetup{justification=centering}
          \caption{Fingerprint robustness on the differentially private Geolife~\cite{zheng2010geolife} dataset (by PIM~\cite{xiao2015protecting}) against a) random flipping attacks, b) correlation-based flipping attacks, c) majority collusion attacks, and d) probabilistic collusion attacks using three methods, i.e., 1) our scheme (represented as "Ours" in the legends), 2) the Boneh-Shaw codes~\cite{boneh1998collusion} ("BS"), and 3) the Tardos codes~\cite{tardos2008optimal} ("TD") for different fingerprinting ratio $p$.}
          
          %with regard to accuracy and utility, i.e., (a) against random distortion attacks on the noisy dataset, (b) against random distortion attacks for different privacy levels $\epsilon$, and (c) against correlated distortion attacks for different privacy levels $\epsilon$.}}
        \label{fig:dp_accuracy}
        \vspace{-2mm}
\end{figure*}

\begin{figure*}[h]
     \centering
    \includegraphics[width=\linewidth]{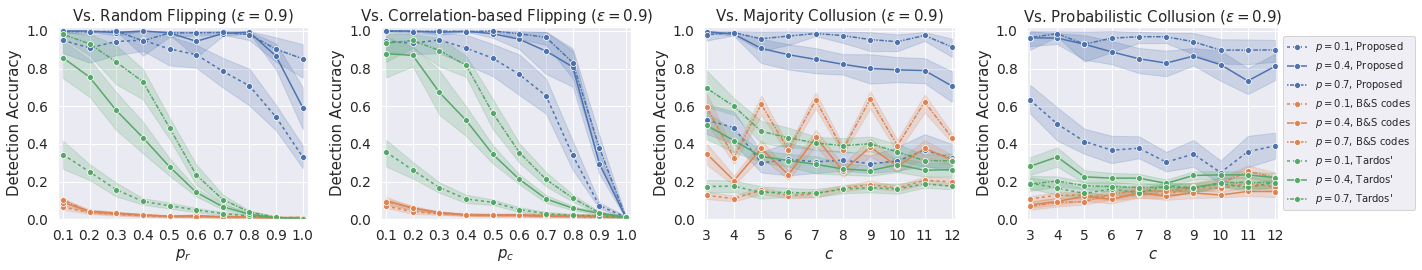}{\hspace{5.7cm}}\\
    \vspace{-0.5cm}
    \hspace{1.15cm}
    \subfloat[vs. RF]{\label{fig:alt_dp_accuracy_a}\hspace{2cm}}
    \hfill
    \subfloat[vs. CF]{\label{fig:alt_dp_accuracy_b}\hspace{2cm}}
    \hfill
    \subfloat[vs. MJR]{\label{fig:alt_dp_accuracy_c}\hspace{2cm}}
    \hfill
    \subfloat[vs. PROB]{\label{fig:alt_dp_accuracy_d}\hspace{2cm}}
    \hspace{2.3cm}
     \vspace{-3mm}
          \captionsetup{justification=centering}
          \caption{Fingerprint robustness on the differentially private Geolife~\cite{zheng2010geolife} dataset protected by an alternative method (AdaTrace~\cite{gursoy2018utility}) against a) random flipping attacks, b) correlation-based flipping attacks, c) majority collusion attacks, and d) probabilistic collusion attacks using three methods, i.e., 1) our scheme (represented as "Ours" in the legends), 2) the Boneh-Shaw codes~\cite{boneh1998collusion} ("BS"), and 3) the Tardos codes~\cite{tardos2008optimal} ("TD") for different fingerprinting ratio $p$.}
          
          %with regard to accuracy and utility, i.e., (a) against random distortion attacks on the noisy dataset, (b) against random distortion attacks for different privacy levels $\epsilon$, and (c) against correlated distortion attacks for different privacy levels $\epsilon$.}}
       \vspace{-3.5mm} \label{fig:alt_dp_accuracy}
\end{figure*}

\begin{table*}[h]
%%\vspace{4mm}
\caption{\textit{Utility Evaluation}. "DSFS" is the proposed scheme in this paper, "BS" denotes the Boneh-Shaw codes, and "Tardos" refers to the Tardos codes. Better results are marked in bold. For Popularity KT coefficient, higher values are better. For the rest metrics, lower is better.}
\vspace{-2mm}
\footnotesize
\centering
\begin{tabular}{cc|ccc|ccc|ccc}
\hline
                             &                 & \multicolumn{3}{c|}{$\epsilon=0.9$}                         & \multicolumn{3}{c|}{$\epsilon=1.7$}                 & \multicolumn{3}{c}{$\epsilon=2.5$}                        \\
                             &                 & DSFS                  & BS~\cite{boneh1998collusion}                  & Tardos~\cite{tardos2008optimal}        & DSFS                  & BS~\cite{boneh1998collusion}           & Tardos~\cite{tardos2008optimal}       & DSFS                 & BS~\cite{boneh1998collusion}                  & Tardos~\cite{tardos2008optimal}       \\ \hline
\multirow{6}{*}{GeoLife~\cite{zheng2010geolife}}     & QA Area AvRE    & \textbf{9.6 $\pm$ 3.6}   & 12.2 $\pm$ 3.3         & 18.7 $\pm$ 5.3   & \textbf{2.8 $\pm$ 0.9}   & 3.9 $\pm$ 1.6   & 3.3 $\pm$ 1.4   & \textbf{0.9 $\pm$ 0.4}  & 1.6 $\pm$ 0.7          & 1.3 $\pm$ 0.3   \\
                             & QA Pattern AvRE & \textbf{2.5 $\pm$ 0.4}   & 4.3 $\pm$ 0.5         & 4.8 $\pm$ 0.5    & \textbf{1.0 $\pm$ 0.2}   & 2.0 $\pm$ 0.3   & 1.8 $\pm$ 0.4   & \textbf{0.5 $\pm$ 0.1}  & 1.0 $\pm$ 0.2          & 0.9 $\pm$ 0.2   \\
                             & Popularity KT~\cite{gursoy2018utility}   & \textbf{0.62 $\pm$ 0.01}  & 0.56 $\pm$ 0.01          & 0.57 $\pm$ 0.02    & \textbf{0.74 $\pm$ 0.01}   & 0.68 $\pm$ 0.02   & 0.69 $\pm$ 0.01   & \textbf{0.83 $\pm$ 0.02}  & 0.79 $\pm$ 0.02          & 0.78 $\pm$ 0.01   \\
                             & Trip Error~\cite{gursoy2018utility}      & \textbf{0.75 $\pm$ 0.01}   & 0.81 $\pm$ 0.01          & 0.81 $\pm$ 0.01    & \textbf{0.66 $\pm$ 0.01}   & 0.78 $\pm$ 0.01   & 0.79 $\pm$ 0.01   & \textbf{0.54 $\pm$ 0.02}  & 0.71 $\pm$ 0.01          & 0.69 $\pm$ 0.01   \\
                             & Diameter Error~\cite{gursoy2018utility}  & \textbf{0.14 $\pm$ 0.00 }  & 0.31 $\pm$ 0.00          & 0.31 $\pm$ 0.00    & \textbf{0.12 $\pm$ 0.00}  & 0.24 $\pm$ 0.00   & 0.24 $\pm$ 0.00   & \textbf{0.11 $\pm$ 0.00}  & 0.21 $\pm$ 0.00          & 0.20 $\pm$ 0.00   \\
                             & DTW Distance    & \textbf{308 $\pm$ 10} & 409 $\pm$ 9        & 400 $\pm$ 11 & \textbf{146 $\pm$ 5} & 182 $\pm$ 6 & 180 $\pm$ 7 & \textbf{78 $\pm$ 3} & 101 $\pm$ 3        & 101 $\pm$ 4 \\ \hline
\multirow{6}{*}{Taxi~\cite{moreira2013predicting}}        & QA Area AvRE    & \textbf{8.4 $\pm$ 3.5}   & 9.6 $\pm$ 4.0          & 13.7 $\pm$ 5.1   & \textbf{0.7 $\pm$ 0.4}  & 2.1 $\pm$ 1.4   & 1.8 $\pm$ 1.3   & 0.34 $\pm$ 0.25           & \textbf{0.33 $\pm$ 0.23} & 0.53 $\pm$ 0.29   \\
                             & QA Pattern AvRE & \textbf{7.5 $\pm$ 2.0}   & 9.3 $\pm$ 1.7          & 9.2 $\pm$ 1.4    & \textbf{0.85 $\pm$ 0.44}  & 1.83 $\pm$ 0.69   & 2.99 $\pm$ 1.39   & \textbf{0.25 $\pm$ 0.07}  & 0.74 $\pm$ 0.22          & 0.66 $\pm$ 0.20   \\
                             & Popularity KT~\cite{gursoy2018utility}   & \textbf{0.54 $\pm$ 0.03}            & 0.53 $\pm$ 0.02 & 0.51 $\pm$ 0.03    & \textbf{0.69 $\pm$ 0.04}  & 0.68 $\pm$ 0.03   & 0.68 $\pm$ 0.02   & \textbf{0.83 $\pm$ 0.05}  & 0.80 $\pm$ 0.03          & 0.77 $\pm$ 0.04   \\
                             & Trip Error~\cite{gursoy2018utility}      & \textbf{0.69 $\pm$ 0.01}   & 0.80 $\pm$ 0.01          & 0.80 $\pm$ 0.01    & \textbf{0.45 $\pm$ 0.02}   & 0.77 $\pm$ 0.01   & 0.77 $\pm$ 0.01   & \textbf{0.36 $\pm$ 0.02}  & 0.64 $\pm$ 0.01          & 0.63 $\pm$ 0.02   \\
                             & Diameter Error~\cite{gursoy2018utility}  & \textbf{0.11 $\pm$ 0.00 }  & 0.30 $\pm$ 0.00          & 0.29 $\pm$ 0.00    & \textbf{0.07 $\pm$ 0.00}   & 0.21 $\pm$ 0.00   & 0.21 $\pm$ 0.00   & \textbf{0.06 $\pm$ 0.00}  & 0.17 $\pm$ 0.00          & 0.17 $\pm$ 0.00   \\
                             & DTW Distance    & \textbf{196 $\pm$ 4} & 257 $\pm$ 6       & 249 $\pm$ 5  & \textbf{75 $\pm$ 3} & 98 $\pm$ 2  & 100 $\pm$ 3  & \textbf{42 $\pm$ 1} & 54 $\pm$ 2         & 56 $\pm$ 2  \\ \hline
\multirow{6}{*}{Oldenburg~\cite{brinkhoff2002framework}}   & QA Area AvRE    & \textbf{1.4 $\pm$ 0.3}   & 2.0 $\pm$ 0.6         & 2.7 $\pm$ 0.6    & \textbf{0.34 $\pm$ 0.13}   & 0.37 $\pm$ 0.10   & 0.41 $\pm$ 0.11   & \textbf{0.13 $\pm$ 0.04}  & 0.18 $\pm$ 0.07          & 0.16 $\pm$ 0.07   \\
                             & QA Pattern AvRE & \textbf{3.4 $\pm$ 0.5}   & 6.3 $\pm$ 0.3          & 6.5 $\pm$ 0.5   & \textbf{1.6 $\pm$ 0.2}   & 3.3 $\pm$ 0.2   & 3.0 $\pm$ 0.2   & \textbf{0.8 $\pm$ 0.1}  & 2.0 $\pm$ 0.2          & 1.9 $\pm$ 0.1   \\
                             & Popularity KT~\cite{gursoy2018utility}   & 0.69 $\pm$ 0.01            & \textbf{0.70 $\pm$ 0.01} & \textbf{0.70 $\pm$ 0.01}    & \textbf{0.84 $\pm$ 0.01}   & 0.83 $\pm$ 0.01   & 0.83 $\pm$ 0.01   & \textbf{0.90 $\pm$ 0.01}  & 0.89 $\pm$ 0.01          & 0.89 $\pm$ 0.01   \\
                             & Trip Error~\cite{gursoy2018utility}      & \textbf{0.70 $\pm$ 0.01}   & 0.80 $\pm$ 0.01          & 0.80 $\pm$ 0.01    & \textbf{0.53 $\pm$ 0.02}   & 0.76 $\pm$ 0.01   & 0.76 $\pm$ 0.01   & \textbf{0.44 $\pm$ 0.02}  & 0.67 $\pm$ 0.02          & 0.66 $\pm$ 0.01   \\
                             & Diameter Error~\cite{gursoy2018utility}  & \textbf{0.11 $\pm$ 0.00}   & 0.28 $\pm$ 0.00          & 0.28 $\pm$ 0.00    & \textbf{0.08 $\pm$ 0.00}   & 0.19 $\pm$ 0.00   & 0.19 $\pm$ 0.00   & \textbf{0.07 $\pm$ 0.00}  & 0.15 $\pm$ 0.00          & 0.14 $\pm$ 0.00   \\
                             & DTW Distance    & \textbf{234 $\pm$ 6} & 264 $\pm$ 7        & 265 $\pm$ 6  & \textbf{86 $\pm$ 3}  & 97 $\pm$ 1  & 96 $\pm$ 2  & \textbf{48 $\pm$ 1} & 55 $\pm$ 1        & 56 $\pm$ 1  \\ \hline
\multirow{6}{*}{San Joaquin~\cite{brinkhoff2002framework}} & QA Area AvRE    & \textbf{2.0 $\pm$ 0.6}   & 2.3 $\pm$ 0.6          & 2.12 $\pm$ 0.60    & \textbf{0.4 $\pm$ 0.1}   & 0.5 $\pm$ 0.2   & 0.6 $\pm$ 0.2   & 0.17 $\pm$ 0.07           & \textbf{0.14 $\pm$ 0.04} & 0.21 $\pm$ 0.06   \\
                             & QA Pattern AvRE & \textbf{3.4 $\pm$ 0.5 }  & 6.6 $\pm$ 0.5          & 6.2 $\pm$ 0.3    & \textbf{1.1 $\pm$ 0.2}   & 3.1 $\pm$ 0.3   & 2.8 $\pm$ 0.4   & \textbf{0.7 $\pm$ 0.1}  & 1.7 $\pm$ 0.2          & 1.5 $\pm$ 0.1   \\
                             & Popularity KT~\cite{gursoy2018utility}   & \textbf{0.68 $\pm$ 0.01}   & 0.65 $\pm$ 0.02          & 0.65 $\pm$ 0.02    & \textbf{0.81 $\pm$ 0.01}   & 0.80 $\pm$ 0.01   & 0.79 $\pm$ 0.01   & \textbf{0.89 $\pm$ 0.01}  & 0.87 $\pm$ 0.01          & 0.87 $\pm$ 0.01   \\
                             & Trip Error~\cite{gursoy2018utility}      & \textbf{0.65 $\pm$ 0.02}   & 0.81 $\pm$ 0.01          & 0.81 $\pm$ 0.01    & \textbf{0.47 $\pm$ 0.01}   & 0.75 $\pm$ 0.01   & 0.76 $\pm$ 0.01   & \textbf{0.39 $\pm$ 0.02}  & 0.66 $\pm$ 0.01          & 0.65 $\pm$ 0.01   \\
                             & Diameter Error~\cite{gursoy2018utility}  & \textbf{0.09 $\pm$ 0.00}   & 0.28 $\pm$ 0.00          & 0.28 $\pm$ 0.00    & \textbf{0.07 $\pm$ 0.00}   & 0.18 $\pm$ 0.00   & 0.18 $\pm$ 0.00   & \textbf{0.05 $\pm$ 0.00}  & 0.14 $\pm$ 0.00          & 0.14 $\pm$ 0.00   \\
                             & DTW Distance    & \textbf{238 $\pm$ 8} & 277 $\pm$ 6        & 271 $\pm$ 5  & \textbf{97 $\pm$ 2}  & 107 $\pm$ 3 & 110 $\pm$ 4 & \textbf{53 $\pm$ 2} & 60 $\pm$ 1         & 59 $\pm$ 2  \\ \hline
\end{tabular}
\vspace{-4mm}
\label{tab:utility_table}
\end{table*}

\vspace{-1.5mm}
\subsection{Fingerprint Robustness}
\label{sec:fp_robustness}
We show the experiment results of fingerprint robustness against five attacks in Section~\ref{sec:threats}, i.e., random flipping attacks, correlation-based flipping attacks, majority collusion attacks, probabilistic collusion attacks, and re-fingerprinting attacks. Here, we represent four of the attacks using abbreviation for simplicity. In particular, "RF" denotes random flipping attacks and "CF" represents correlation-based flipping attacks. "MJR" and "PROB" are majority collusion attacks and probabilistic collusion attacks, respectively. 

\subsubsection{Fingerprint Robustness on Datasets Without Differential Privacy.}
\label{sec:raw_accuracy}
We first evaluate fingerprint robustness of our proposed scheme without differential privacy. Figure~\ref{fig:raw_accuracy} shows the performance of the proposed scheme against multiple attacks. 
For random flipping attacks, our scheme achieves almost $100\%$ accuracy if the attacker does not perturb more than $60\%$ of the location points, and it decreases to $90\%$ if the attacker distorts $80\%$ of the location points. In terms of correlation-based flipping attacks, the scheme has high accuracy when the flipping ratio $p_c$ is less or equal to $0.6$, and the accuracy drops significantly for larger $p_c$. The reason is almost the same as why the probabilistic fingerprinting scheme (PFS)~\cite{yilmaz2020collusion} does not work on location datasets, i.e., the forced deviation (shown in Figure~\ref{fig:deathloop}). For an acceptable data utility, the attacker does not prefer a large $p_c$ in practice. For majority collusion attacks, the detection accuracy of our scheme is more than $80\%$. In terms of probabilistic collusion attacks, our scheme achieves $99\%$ detection accuracy if $c=3$ and still get around $60\%$ if $c$ increases to $12$. 

Note that the scheme does not benefit from higher fingerprinting ratio against two correlation-based attacks (i.e., correlation-based flipping attacks and probabilistic collusion attacks) for non-noisy datasets, and the accuracy becomes even worse for probabilistic collusion attacks, which can be explained as follows. In a non-noisy trajectory, pairwise correlations mostly hold, i.e., the transition probability from the previous point to the current point remains high. In that case, if we fingerprint (modify) two consecutive points, pairwise correlation between the modified values mostly decrease. When fingerprinting ratio is high (i.e., $p>0.5$), such scenarios occur more commonly and they can be exploited by the attacker, resulting in a decrease in detection accuracy. 
On the other hand, when the fingerprinting ratio is low, e.g., $p=0.1$, our scheme does not have too few fingerprinted points to provide fingerprint robustness against the collusion attacks, resulting in a degradation in accuracy.
For optimal performance of our scheme, we recommend using a fingerprinting ratio around $0.4$. This ensures high data utility and high fingerprint robustness in the shared dataset simultaneously.

In comparison to our scheme, the two existing methods are not equally robust. The Boneh-Shaw codes achieve around $50\%$ detection accuracy in majority collusion attacks and have at most $20\%$ chance to identify the attacker against other attacks, where the wavy style in Figure~\ref{fig:raw_accuracy}c results from its own design. Detection accuracy of the Tardos codes is $100\%$ against the two flipping attacks if flipping ratio is $0.1$, but it quickly drops to below $40\%$ and $20\%$ for random flipping attacks and correlation-based flipping attacks, respectively. For collusion attacks, their detection accuracy are at most $70\%$ if $3$ attackers collude and $40\%$ when $12$ are involved. Overall, our scheme achieves better performance against all the considered attacks.

\vspace{-1.5mm}
\subsubsection{Fingerprint Robustness on Differentially Private Datasets Using PIM}
\label{sec:dp_accuracy}
For differentially private datasets, our scheme performs significantly better (shown in Figure~\ref{fig:dp_accuracy}). The proposed scheme achieves around $99.9\%$ detection accuracy against random flipping attacks, majority collusion attacks (with a little drop for larger collusion count), and probabilistic collusion attacks. In terms of correlation-based flipping attacks, our scheme achieves $99.9\%$ accuracy when $p_c<0.8$, and it achieves $98\%$ accuracy if $p_c = 0.8$. As introduced in Section~\ref{sec:raw_accuracy}, higher $p_c$ is not desired by the attacker since correlations in the dataset barely holds. In conclusion, this implies that the post-processing mechanism that is integrated in our proposed fingerprinting scheme actually works well, and the detection benefits from it on the datasets with lower correlations (e.g., a differentially private dataset). Meanwhile, the Boneh-Shaw codes and the Tardos codes have similar performance to the ones in Section~\ref{sec:raw_accuracy}, and our scheme outperforms those methods.

\begin{figure}[h]
 \centering
  \includegraphics[width=0.28\textwidth]{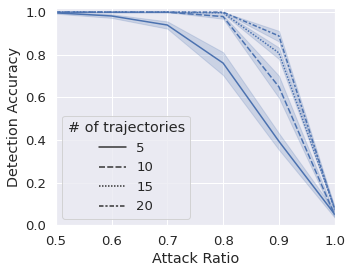}
  \caption{Fingerprint robustness against re-fingerprinting attacks on differentially private datasets of different sizes. The attack ratio denotes the fingerprinting ratio that the attacker uses during the attack. }
\label{fig:fp_attack}
\end{figure}

\vspace{-1.5mm}
\subsubsection{Fingerprint Robustness on Differentially Private Datasets Using Other Methods}
\label{sec:eval_alt}
In order to show our framework works with different differentially private mechanisms on location datasets, we implement an alternative DP mechanism, i.e., AdaTrace~\cite{gursoy2018utility} instead of PIM. 
However, AdaTrace is a synthetic mechanism that does not preserve any temporal information in the released dataset. Thus, we post-process the output dataset from AdaTrace using the Bresenham's algorithm, a line drawing algorithm, to traverse all passed points between two points and add time-sequenced indexes to each point. 
By using the Bresenham's algorithm and then assigning timestamps manually, we generate a synthetic dataset with high pair-wise correlations but a fake version. 
As shown in Figure~\ref{fig:alt_dp_accuracy}, we achieve similar results compared with the ones on original datasets  (i.e., in Figure~\ref{fig:raw_accuracy}) in general, which proves that our scheme can work on other differentially private mechanisms.

\vspace{-1.5mm}
\subsubsection{Fingerprint robustness against re-fingerprinting attack}
\label{sec:eval_fp_attack}
The attacker can distort the embedded fingerprint by applying the proposed fingerprinting scheme on the received dataset, namely re-fingerprinting attacks. To evaluate our scheme against such attacks, we design the experiment as follows. We build small datasets of different sizes for the evaluation. We assume that the attacker, based on the experiment results in Sections~\ref{sec:raw_accuracy},~\ref{sec:dp_accuracy}, and~\ref{sec:eval_alt}, chooses the optimal parameters, i.e., $\tau=0.005, \theta=0.5$, and applies the proposed fingerprinting scheme to the received dataset (which is also fingerprinted by the data owner). The attack ratio in this attack refers to the fingerprinting ratio that the attacker uses. As shown in Figure~\ref{fig:fp_attack}, our scheme offers high fingerprint robustness against re-fingerprinting attacks regardless of the number of trajectories in the dataset, while larger datasets (with more trajectories) lead to higher detection accuracy. When the trajectory count exceeds $10$, the detection accuracy achieves $98\%$ for any fingerprinting ratio no larger than $0.8$ and still stays above $60\%$ even if the attack ratio reaches to $0.9$. For higher attack ratios, similar to correlation-based attacks discussed in Section~\ref{sec:raw_accuracy}, the resulting low data utility limits the attacker from executing such attacks.

\vspace{-1mm}
\subsection{Utility Evaluation}
Table~\ref{tab:utility_table} shows the data utility of the proposed scheme and compares it with the original dataset. For $\epsilon = 0.9, 1.7, \text{ and } 2.5$, our proposed method is better than the Boneh-Shaw codes and Tardos codes in most cases. Meanwhile, our scheme is not the best for query answering (when $\epsilon = 2.5$) on the Taxi and San Joaquin datasets and for popularity analysis (when $\epsilon = 0.9$) on the Oldenburg dataset. On the Taxi dataset when $\epsilon=2.5$, the error for query answering on location points is $0.34$, which is slightly higher than $0.33$ for the Boneh-Shaw codes. Similarly, the performance of our scheme when $\epsilon=0.9$ is $0.01$ worse than the two existing methods for popularity analysis on the Oldenburg dataset. On the San Joaquin dataset, the query answering error on location points of our scheme
is $0.17$, while the error is $0.14$ if the Boneh-Shaw codes are used.
For the majority of the metrics, our scheme outperforms the Boneh-Shaw codes and the Tardos' code. For a few metrics, our scheme is slightly worse but still comparable with the existing methods.

% provides better correlation fit rate than other ones, and we provide the experiment results in Appendix \ref{sec:cfr}. Comparing the output from PIM and from the whole system (including PIM and HPFS), we can conclude that our fingerprinting scheme provides high data utility for all metrics. 
%\vspace{-5mm}
\subsubsection{Computation Time}
\label{eval:time}
We present the computation time of the proposed scheme in Table~\ref{table:time_cost}. For a dataset of $100$ trajectories with length equal to $500$, the proposed scheme only takes $2.5779$ seconds to generate one fingerprinted copy. We observe that the computation time increases linearly with the increasing trajectory length. In conclusion, our scheme shows practical time efficiency for fingerprint generation and scales well for large datasets.

\begin{table}[h]
%\vspace{4mm}
\footnotesize
\centering
\caption{Execution time of generating a fingerprinted dataset ($n = 100$)}
\vspace{-3mm}
\begin{tabular}{ c c c c c c c c c c c }
\hline
$l$ & 100 & 200 & 300 & 400 & 500 %& 600 & 700 & 800 & 900 & 1,000
\\
\hline
time$(s)$ & 0.5177 & 1.0229 & 1.5268 & 2.0407 & 2.5779 %& 3.0555 & 3.4493 &  4.0604& 4.5701 & 5.0802 
\\
\hline
\end{tabular}
\vspace{-4.5mm}
\label{table:time_cost}
\end{table}

\section{Discussion}
\label{sec:disc}
\vspace{-1mm}
Here, we compare PIM with other differentially private mechanisms and discuss the correlation model.
\vspace{-1.5mm}
\subsection{Comparison Between PIM and Other Differentially Private Mechanisms}
\label{sec:comparison}

Compared with PIM, other existing works more or less have their limitations for realistic location dataset sharing.  %and cannot be used for general sharing of location trajectories. 
Jiang et al.'s approach~\cite{jiang2013publishing} requires that the starting and finishing points of all the trajectories should be fixed, making it only work on specific types such as ship or flight trajectories. \cite{gursoy2018utility} and~\cite{he2015dpt} need accurate statistical features from the input dataset. Thus, the size of the dataset should be comparably large. In other words, they cannot handle datasets with only a few trajectories. In addition, trajectory addition and removal is one of the most common requests from the users as they become more concerned about their data privacy~\cite{bourtoule2021machine}.
Synthetic methods~\cite{gursoy2018utility, he2015dpt} cannot perform such operations by simply working on the protected dataset and they have to regenerate the entire dataset. Meanwhile, PIM is executed on each trajectory instead of the whole dataset. It can easily achieve this by adding or removing generated copies of a specific trajectory to/from the shared dataset.
% pA two post-processing schemes, i.e., the post-processing scheme and  with one Thus, we claim that both the post-processing scheme so using public correlations are not considered as sume  iThe scheme modifies each location point only based on auxiliary information, i.e., the correlation model, without considering the true values of the data points before the differentially private generation. Thus, the post-processing scheme satisfies the property of immunity to post-processing and does not violate the DP guarantee of PIM.

% PIM provides differential privacy, and the following utility-focused post-processing and fingerprinting scheme does not violate the differential privacy guarantee. In the beginning, PIM offers initial guarantee differential privacy for each trajectory in the dataset~\cite{xiao2015protecting}.  Thus, the utility-focused post-processing does not break differential privacy since it only utilizes the noisy trajectories generated from PIM and the auxiliary information (i.e., correlations) from public sources. Meanwhile, the fingerprinting scheme also preserves the guarantee, which can be proved using the similar logic as above since it does not utilize any information from the original dataset. Therefore, we prove that our entire approach satisfies differential privacy.

\subsection{Correlation Model}
\label{sec:corr_discussion}
In this work, we use $2$-gram Markov chain to model correlations. If we use a higher-order model, each pattern $X$'s occurrence will decrease significantly since longer prefixes are harder to find intuitively. Therefore, we cannot collect enough patterns $Xg$ to form a reliable transition distribution for a prefix $X$, thus resulting in an inaccurate transition matrix. 
For instance, GeoLife dataset~\cite{zheng2010geolife} consists of $17,621$ trajectories in Beijing. However, we can hardly construct a reliable $3$-gram model out of it, especially if we use a dense grid for services like Google Maps that collects location data frequently. Some approaches use a sparse grid to overcome this problem~\cite{xiao2015protecting, chen2012differentially} (around $400 * 400 m^2$), but the location points are too general for analytical purposes. On the other hand, our target applications, e.g., Google Maps and outdoor exercises, cannot bear such general locations. As a result, we compromise with the $2$-gram Markov chain.

\vspace{-3mm}
\section{Conclusion and Future Work}
\label{sec:conclusion}
In this paper, we design a system that achieves both privacy preservation and robust fingerprinting for location datasets. We first apply a differentially private mechanism to the dataset and then implement a fingerprinting scheme that considers pairwise correlations in the location data and prevents the attackers from unauthorized leakage of the dataset. With the integration of a utility-boosting post-processing, our proposed direction-sensitive fingerprinting scheme provides high data utility for data analyzers.

There are several directions for further research. First, we plan to improve our correlation model to a higher-order model (e.g., using road structures) and analyze the performance of the scheme. %Using a non-uniform grid in discretization %such as based on population density is also an interesting topic.
In addition, a non-uniform grid in discretization can be used and different types of collusion attacks can be defined and studied. Moreover, our approach provides differential privacy and fingerprint robustness in two separate steps. Combining those two steps
is another potential future work.
% First, the differential privcay mechanism (PIM) we use in the evaluation provides location point-level privacy guarantee, as we cannot find other existing approach suitable for single trajectory release. If a better DP method, e.g., providing quantifiable privacy on the whole trajectory, appears in the future, we can deploy deeper analysis regarding privacy loss and more.
% Second, we do not cover all possible attacks in this paper. For example, subset attacks along with adding noises to timestamps are powerful attacks since the detection mechanism needs far more efforts in aligning trajectories, which becomes worse if the adversary does not report timestamps at all. Other colluding attacks such as average colluding attacks are also interesting topics for the next steps.

%%% Local Variables:
%%% mode: latex
%%% TeX-master: "main"
%%% End:

\bibliographystyle{ACM-Reference-Format}
\bibliography{bib}

% % --- Appendix ---%
 \appendix
 \vspace{3mm}
\noindent{\Large{\textbf{Appendices}}}
\appendix
\label{sec:appendix}

\section{Proof of immunity to Post-Processing of Our Scheme}
\label{sec:immunity_discussion}

\begin{clm}
The utility-focused post-processing scheme and the direction-sensitive fingerprinting scheme do not violate differential privacy. 
\begin{proof}
First, we assume that the input data are released under an arbitrary privacy-preserving mechanism that satisfies $\epsilon$-differential privacy. 
According to Proposition~\ref{prop:immunity}, any mapping function does not violate the guarantee of a differentially-private method if the function does not utilize the values from the original dataset. 
During utility-focused post-processing in Section~\ref{sec:postprocess} and direction-sensitive fingerprinting in Section~\ref{sec:DSFS}, we only use public correlations instead of any information from the original dataset, which satisfies the conditions of immunity to post-processing.
Thus, the post-processing and fingerprinting do not violate $\epsilon$-differential privacy that is provided by the privacy-preserving mechanism.

\end{proof}
\end{clm}

\section{Existing Fingerprinting Schemes}
In this section, we introduce the two existing fingerprinting schemes that we use for comparison.

\vspace{-2mm}
\subsection{The Boneh-Shaw Codes}
\label{sec:bs}
The Boneh-Shaw codes~\cite{boneh1998collusion} are collusion-resistant fingerprinting codes. It catches one of the colluding parties with only a probability $\omega$ of incorrect accusation with $c$ service providers colluding ($\omega$-secure) under the marking assumption~\cite{boneh1998collusion}. 
% A codeword set for $4$ users consists of $3$ digits: $\{111, 011, 001, 000\}$, and each user gets a one of them.
$\Gamma(n, d)$-codes serve $n$ users and consist of $n*d$ digits. Each user $i \in [1,n]$ gets the first $(i-1)*d$ bits as $1$'s and the rest $(n-i)*d$ as $0$'s. An example of the $\Gamma(4,3)$ code is: \{111-111-111, 000-111-111, 000-000-111, 000-000-000\}, and each user receives one of the codewords. By identifying the first block with a majority of $1$'s in the leaked data, e.g., the $i$-th block, the algorithm considers the user $i$ as guilty. In the above example, if 001-011-111 is leaked, the $2$nd user who owns 000-111-111 will be accused of leaking the data since the $2$nd block is the first block with a majority of $1$'s.

\vspace{-2mm}
\subsection{The Tardos Codes}
\label{sec:tardos}
The Tardos codes~\cite{tardos2008optimal} are another binary fingerprinting technique under the marking assumption. The codes utilize randomization in construction and provide similar security against majority collusion attacks while requiring a shorter code length than the Boneh-Shaw codes. % in the same setting.
%$\omega$-secure against colluding attacks, while they require a shorter code length than the Boneh-Shaw codes in the same setting.
The construction of the Tardos codes requires the number of sharings $n$, the number of colluding units $c$, and the expected security $\omega$. The minimal binary code length to ensure $\omega$-security is $m = 100c^2k$, where $k=\lceil \log (1/\omega)\rceil$. % ($\log$ refers to natural logarithm). 
Let $t = 1/(300c)$ and $sin^2t'=t, 0<t'<\pi /4$. $p_i$ denotes the probability of $1$ at position $i$, i.e., $Pr(X_i=1)=p_i$, and is independently calculated. To select the probability for each position $i$, we sample $r_i \in [t', \pi /2-t']$ uniformly and then acquire $p_i = sin^2r_i$. Let $X_{ji}$ denotes the $i$-th digit of the user $j$ and $\mathcal{Y}=\{y_1, y_2, \dots, y_m\}$ denotes the leaked data. While accusing the colluders, the codes use a scoring function as 
\begin{equation}
  U_{ji} =
    \begin{cases}
      \sqrt{\frac{1-p_i}{p_i}}  & \text{if $X_{ji} = 1$}\\
      -\sqrt{\frac{p_i}{1-p_i}}  & \text{if $X_{ji} = 0$}\\
    \end{cases}   
    \vspace{-1mm}
\end{equation}
and accuse the user $j$ if $$\sum_{i=1}^m y_i U_{ji} \geq 20ck$$.

\vspace{-3mm}
\section{Decentralized Setting}
\label{sec:decentral_discussion}
We build our system in the centralized setting, i.e., users' location points are collected by a centralized data server (service provider) and then processed by our scheme. This relies on an honest party involved into the system since the centralized data server (i.e., the service provider) has direct access to the collected dataset. If no such party exists, we can alternatively set up a decentralized system, where the privacy is protected before sending location data to the centralized server. In such a decentralized setting, users can apply the DP protection locally on their devices by setting the desired privacy level they want to achieve. Then, the protected data are transmitted to the centralized server. Every time when the service provider collects real-time location information from the users, users instantly protect their locations under differential privacy (e.g., using PIM) and send the noisy locations to the centralized server. The server collects these locations sequentially and apply our proposed fingerprinting scheme to location data. In this case, real locations are not exposed to any party including the centralized server, thus protecting users' location privacy in a better way. However, this setting sacrifices users' experience while using location-based servers, and thus some service providers may offer poor services due to the inaccuracy of the location information. While using Google Maps for navigation, one does not want to report incorrect locations. But if one uses Google Maps to find nearby restaurants, they often accept a vague or slightly deviated localization. Service providers can choose either setting based on the services they provide.

\section{Trajectory Post-Processing Scheme}
\label{app:postprocess}
Algorithm \ref{alg:smoothing} shows the steps of the post-processing scheme described in Section \ref{sec:postprocess}, where $\lVert\cdot\rVert_2$ denotes the $l_2$-norm.

\begin{algorithm}[h] 
\small
 \SetKwInOut{Input}{input}
 \SetKwInOut{Output}{output}
 \Input{Noisy trajectory $\hat{\mathcal{X}}=[\hat{x}_1,\hat{x}_2,\dots,\hat{x}_m]$, location alphabet $\mathcal{G}$, conditional probability in the correlations $Pr(x_j|x_{j-1})$ for any $j \in [1,m]$, probability threshold $\tau$}
 \Output{Smoothed trajectory $\mathcal{X^*}=[x_1^*,x_2^*,\dots,x_m^*]$}
  $x_1^* \leftarrow x_1$
  
  \ForAll{$j \in  \{2,3,\ldots,m\}$}
  {
        $ prob_\tau(x^*_{j-1}) \leftarrow \tau$-probable set of $x^*_{j-1}$\;
        $x_{closest} \leftarrow $ closet point to 
        $\hat{x}_j $ in $ prob_\tau(x^*_{j-1})$\;
       \uIf{$\hat{x}_j \notin prob_\tau(x^*_{j-1})$}{
            \uIf{$\lVert  x^*_{j-1}, \hat{x}_j\rVert_2 \leq \lVert x^*_{j-1}, x_{closest}\rVert_2$}{
                $x^*_j \leftarrow \hat{x}_j$
            }
            \uElse{
                $x^*_j \leftarrow x_{closest}$
            }
        }
        \uElse{
            $x^*_j \leftarrow \hat{x}_j$
        }

  }
 \caption{Trajectory Post-Processing Scheme}
 
 \label{alg:smoothing}
\end{algorithm}

% \section{Direction-Sensitive Fingerprinting Scheme}
% \label{sec:dsfs}

% We provide the Hybrid Probabilistic Fingerprinting Scheme in Algorithm~\ref{alg:dsfs}.

\section{Additional Experimental results}
\label{app:add_results}

We show additional experimental results in this section. First, we evaluate fingerprinting robustness on other datasets apart from GeoLife~\cite{zheng2010geolife}. Then we show how length impact fingerprinting performance. Also, we extend Section~\ref{sec:dp_accuracy} and evaluate our scheme's performance when multiple trajectories are leaked.

\subsection{Fingerprinting Robustness on Other Datasets Under Differential Privacy}
\begin{figure*}[h]

    \centering
    \includegraphics[width=\linewidth]{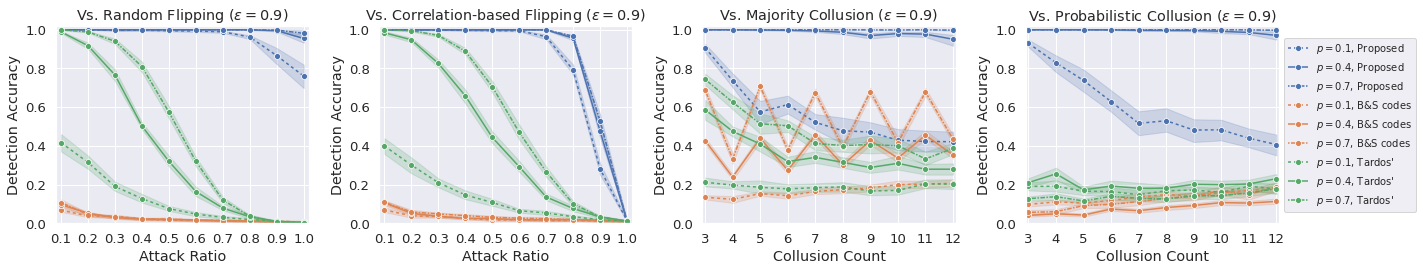}{\hspace{5.7cm}}\\
    \vspace{-0.5cm}
    \hspace{1.15cm}
    \subfloat[vs. RF]{\label{fig:taxi_dp_accuracy_a}\hspace{2cm}}
    \hfill
    \subfloat[vs. CF]{\label{fig:taxi_dp_accuracy_b}\hspace{2cm}}
    \hfill
    \subfloat[vs. MJR]{\label{fig:taxi_dp_accuracy_c}\hspace{2cm}}
    \hfill
    \subfloat[vs. PROB]{\label{fig:taxi_dp_accuracy_d}\hspace{2cm}}
    \hspace{2.3cm}
     \vspace{-3mm}
          \captionsetup{justification=centering}
          \caption{Fingerprint robustness on the differentially private Taxi~\cite{moreira2013predicting} dataset (by PIM~\cite{xiao2015protecting}) against a) random flipping attacks, b) correlation-based flipping attacks, c) majority collusion attacks, and d) probabilistic collusion attacks using three methods, i.e., 1) our scheme (represented as "Ours" in the legends), 2) the Boneh-Shaw codes~\cite{boneh1998collusion} ("BS"), and 3) the Tardos codes~\cite{tardos2008optimal} ("TD") for different fingerprinting ratio $p$.}
          
          %with regard to accuracy and utility, i.e., (a) against random distortion attacks on the noisy dataset, (b) against random distortion attacks for different privacy levels $\epsilon$, and (c) against correlated distortion attacks for different privacy levels $\epsilon$.}}
        \label{fig:taxi_dp_accuracy}
        \vspace{-2mm}
\end{figure*}

\begin{figure*}[h]

    \centering
    \includegraphics[width=\linewidth]{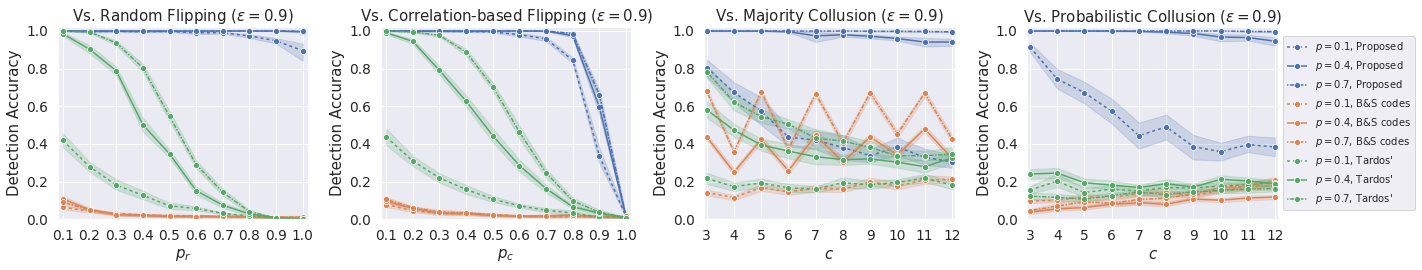}{\hspace{5.7cm}}\\
    \vspace{-0.5cm}
    \hspace{1.15cm}
    \subfloat[vs. RF]{\label{fig:olden_dp_accuracy_a}\hspace{2cm}}
    \hfill
    \subfloat[vs. CF]{\label{fig:olden_dp_accuracy_b}\hspace{2cm}}
    \hfill
    \subfloat[vs. MJR]{\label{fig:olden_dp_accuracy_c}\hspace{2cm}}
    \hfill
    \subfloat[vs. PROB]{\label{fig:olden_dp_accuracy_d}\hspace{2cm}}
    \hspace{2.3cm}
     \vspace{-3mm}
          \captionsetup{justification=centering}
          \caption{Fingerprint robustness on the differentially private OldenBurg~\cite{brinkhoff2002framework} dataset (by PIM~\cite{xiao2015protecting}) against a) random flipping attacks, b) correlation-based flipping attacks, c) majority collusion attacks, and d) probabilistic collusion attacks using three methods, i.e., 1) our scheme (represented as "Ours" in the legends), 2) the Boneh-Shaw codes~\cite{boneh1998collusion} ("BS"), and 3) the Tardos codes~\cite{tardos2008optimal} ("TD") for different fingerprinting ratio $p$.}
          
          %with regard to accuracy and utility, i.e., (a) against random distortion attacks on the noisy dataset, (b) against random distortion attacks for different privacy levels $\epsilon$, and (c) against correlated distortion attacks for different privacy levels $\epsilon$.}}
        \label{fig:olden_dp_accuracy}
        \vspace{-2mm}
\end{figure*}

\begin{figure*}[h]

    \centering
    \includegraphics[width=\linewidth]{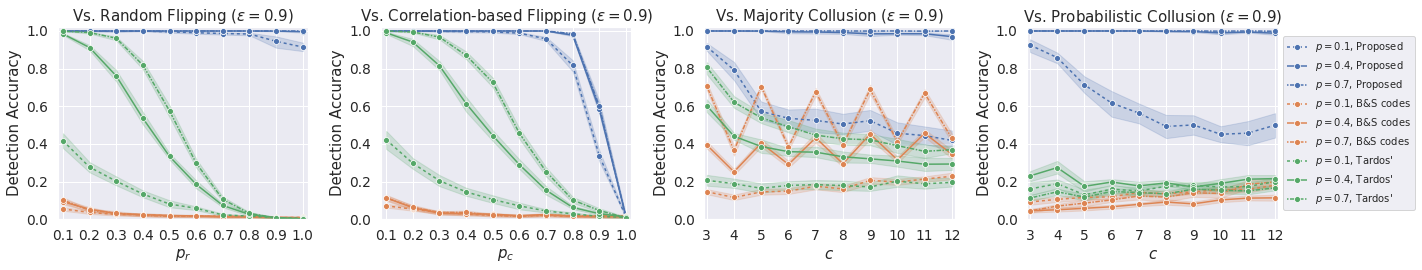}{\hspace{5.7cm}}\\
    \vspace{-0.5cm}
    \hspace{1.15cm}
    \subfloat[vs. RF]{\label{fig:san_dp_accuracy_a}\hspace{2cm}}
    \hfill
    \subfloat[vs. CF]{\label{fig:san_dp_accuracy_b}\hspace{2cm}}
    \hfill
    \subfloat[vs. MJR]{\label{fig:san_dp_accuracy_c}\hspace{2cm}}
    \hfill
    \subfloat[vs. PROB]{\label{fig:san_dp_accuracy_d}\hspace{2cm}}
    \hspace{2.3cm}
     \vspace{-3mm}
          \captionsetup{justification=centering}
          \caption{Fingerprint robustness on the differentially private San Joaquin~\cite{zheng2010geolife} dataset (by PIM~\cite{xiao2015protecting}) against a) random flipping attacks, b) correlation-based flipping attacks, c) majority collusion attacks, and d) probabilistic collusion attacks using three methods, i.e., 1) our scheme (represented as "Ours" in the legends), 2) the Boneh-Shaw codes~\cite{boneh1998collusion} ("BS"), and 3) the Tardos codes~\cite{tardos2008optimal} ("TD") for different fingerprinting ratio $p$.}
          
          %with regard to accuracy and utility, i.e., (a) against random distortion attacks on the noisy dataset, (b) against random distortion attacks for different privacy levels $\epsilon$, and (c) against correlated distortion attacks for different privacy levels $\epsilon$.}}
        \label{fig:san_dp_accuracy}
        \vspace{-2mm}
\end{figure*}

As is shown in Figure~\ref{fig:taxi_dp_accuracy}, \ref{fig:olden_dp_accuracy}, and \ref{fig:san_dp_accuracy}, the results are almost identical to GeoLife~\cite{zheng2010geolife}'s (in Section~\ref{sec:dp_accuracy}. It proves that our fingerprinting scheme is robust and consistent for all location datasets.

\subsection{Fingerprinting Robustness on Differentially Private Datasets for Trajectories with Different Lengths}
\label{eval:length}

\begin{figure*}[h]
     \includegraphics[width=\linewidth]{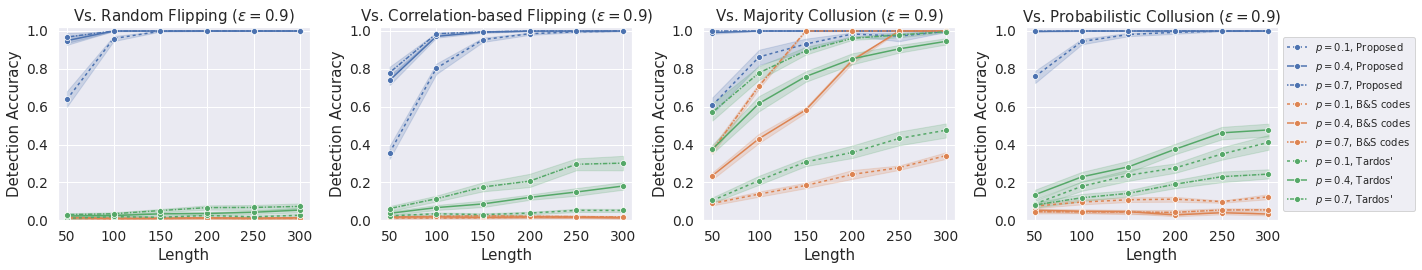}{\hspace{5.7cm}}\\
    \vspace{-0.5cm}
    \hspace{1.15cm}
    \subfloat[vs. RF]{\label{fig:dp_accuracy_length_a}\hspace{2cm}}
    \hfill
    \subfloat[vs. CF]{\label{fig:dp_accuracy_length_b}\hspace{2cm}}
    \hfill
    \subfloat[vs. MJR]{\label{fig:dp_accuracy_length_c}\hspace{2cm}}
    \hfill
    \subfloat[vs. PROB]{\label{fig:dp_accuracy_length_d}\hspace{2cm}}
    \hspace{2.3cm}
     \vspace{-3mm}
          \captionsetup{justification=centering}
          \caption{Fingerprint robustness on differentially private datasets protected (by PIM~\cite{xiao2015protecting}) with different lengths of the leaked trajectory against a) random flipping attacks, b) correlation-based flipping attacks, c) majority collusion attacks, and d) probabilistic collusion attacks using three methods, i.e., 1) our scheme (represented as "Ours" in the legends), 2) the Boneh-Shaw codes~\cite{boneh1998collusion} ("BS"), and 3) the Tardos codes~\cite{tardos2008optimal} ("TD") for different fingerprinting ratios.}
          
        \label{fig:dp_accuracy_length}
\end{figure*}

Figure \ref{fig:dp_accuracy_length} shows fingerprinting robustness on differentially private datasets for trajectories with different lengths.

\subsection{Fingerprinting Robustness on Differentially Private Datasets While Multiple Trajectories are Leaked}
\label{eval:multi_leakage}

\begin{figure*}[h]
     \includegraphics[width=\linewidth]{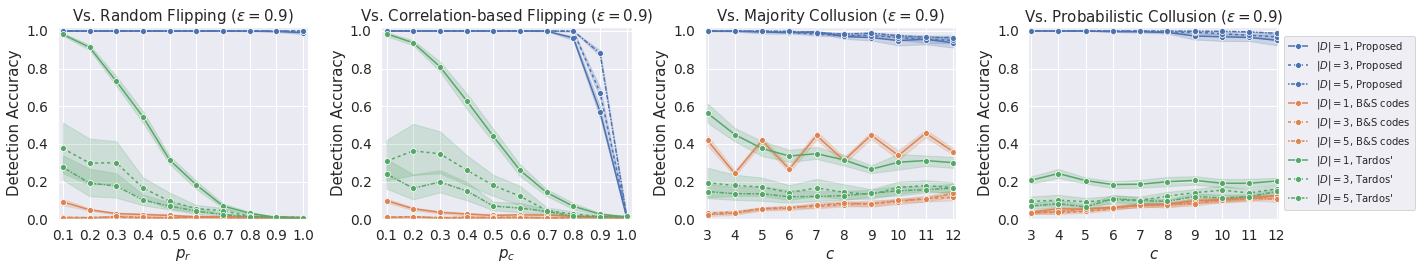}{\hspace{5.7cm}}\\
    \vspace{-0.5cm}
    \hspace{1.15cm}
    \subfloat[vs. RF]{\label{fig:dp_accuracy_multi_leaked_a}\hspace{2cm}}
    \hfill
    \subfloat[vs. CF]{\label{fig:dp_accuracy_multi_leaked_b}\hspace{2cm}}
    \hfill
    \subfloat[vs. MJR]{\label{fig:dp_accuracy_multi_leaked_c}\hspace{2cm}}
    \hfill
    \subfloat[vs. PROB]{\label{dp_accuracy_multi_leaked_d}\hspace{2cm}}
    \hspace{2.3cm}
     \vspace{-3mm}
          \captionsetup{justification=centering}
          \caption{Fingerprint robustness on differentially private datasets protected (by PIM~\cite{xiao2015protecting}) with different number of trajectories in the leaked dataset against a) random flipping attacks, b) correlation-based flipping attacks, c) majority collusion attacks, and d) probabilistic collusion attacks using three methods, i.e., 1) our scheme (represented as "Ours" in the legends), 2) the Boneh-Shaw codes~\cite{boneh1998collusion} ("BS"), and 3) the Tardos codes~\cite{tardos2008optimal} ("TD") for different fingerprinting ratios.}
          
        \label{fig:dp_accuracy_multi_leaked}
\end{figure*}

Figure \ref{fig:dp_accuracy_multi_leaked} shows fingerprinting robustness on differentially private datasets while multiple trajectories are leaked.

\end{document}